\documentclass[alpha-refs]{nbdt-article}
\usepackage{siunitx}
\usepackage{natbib}
\papertype{Original Article}
\paperfield{}

\title{Circumstantial evidence and explanatory models for synapses in large-scale spike recordings}

\author[1-3]{Ian H. Stevenson}

\affil[1]{Department of Psychological Sciences, University of Connecticut}
\affil[2]{Department of Biomedical Engineering, University of Connecticut}
\affil[3]{Connecticut Institute for Brain and Cognitive Science, University of Connecticut}

\corremail{ian.stevenson@uconn.edu}

\fundinginfo{This material is based upon work supported by the National Science Foundation under Grant 1651396.}

\runningauthor{Stevenson}

\begin{document}

\maketitle
\begin{abstract}
Whether, when, and how causal interactions between neurons can be meaningfully studied from observations of neural activity alone are vital questions in neural data analysis. Here we aim to better outline the concept of functional connectivity for the specific situation where systems neuroscientists aim to study synapses using spike train recordings. In some cases, cross-correlations between the spikes of two neurons are such that, although we may not be able to say that a relationship is causal without experimental manipulations, models based on synaptic connections provide precise explanations of the data. Additionally, there is often strong circumstantial evidence that pairs of neurons are monosynaptically connected. Here we illustrate how circumstantial evidence for or against synapses can be systematically assessed and show how models of synaptic effects can provide testable predictions for pair-wise spike statistics. We use case studies from large-scale multi-electrode spike recordings to illustrate key points and to demonstrate how modeling synaptic effects using large-scale spike recordings opens a wide range of data analytic questions. 
\end{abstract}

\section{Introduction}
In many cases, just by chance, large-scale spike recordings will happen to sample pairs of neurons that are monosynaptically connected. The mouse (mus musculus) brain, for instance, has $\sim$ 71 million neurons with $\sim$8 thousand synapses per neuron \citep{Braitenberg_Schuz_1998,Herculano-Houzel_Mota_Lent_2006}. Assuming that neurons are uniformly randomly connected, this would imply that 1 out of every $\sim$9 thousand pairs of neurons are monosynaptically connected. Although this is a small proportion, in a large-scale recording of 1000 neurons ($\sim$1 million possible connections) one could expect to observe $\sim100 \pm10$ monosynaptically connected pairs just by chance (mean $\pm$ s.d.). Since electrode arrays often record from a single brain region or a few adjacent brain regions targeted for their connectivity, connection probabilities are typically much higher in practice. Estimates of connection probabilities from intracellular recordings in cortical slices suggest that for an electrode spacing <100$\mu m$ the chances of two neurons being connected are on the order of 1 in 10 \citep{Holmgren_Harkany_Svennenfors_Zilberter_2003,Seeman_et_al._2018,Song_Sjostrom_Reigl_Nelson_Chklovskii_2005}. Although the number of strongly connected pairs in a typical large-scale spike recording may be small relative to the number of possible pairs, sampling at least some connected pairs will become inevitable as the number of neurons that can be simultaneously recorded increases \citep{Buzsáki_2004,Stevenson_Kording_2011}. Accurately detecting these connections from sparse spiking activity alone is a statistical challenge, but identifying connected pairs can allow us to characterize neural circuits in action.

Evidence for synaptic connections can, in many cases, appear in the cross-correlation between the spike trains of two neurons \citep{Fetz_Toyama_Smith_1991,Gerstein_Perkel_1969,Moore_Segundo_Perkel_Levitan_1970,Perkel_Gerstein_Moore_1967}. For a strong monosynaptic connection, the spiking of the presynaptic neuron affects the probability of postsynaptic spiking in the short time period following presynaptic spikes. All things being equal, an excitatory synaptic connection will increase the postsynaptic spiking probability, while an inhibitory synaptic connection will decrease the postsynaptic spiking probability, relative to the postsynaptic neuron’s baseline firing rate. \cite{Levick_Cleland_Dubin_1972} introduced two key metrics for assessing spike transmission: efficacy and contribution. The synaptic efficacy measures the excess probability of the postsynaptic neuron spiking following a presynaptic spike, while contribution measures the proportion of postsynaptic spikes considered to be the result of presynaptic spikes. For excitatory synapses, efficacy and contribution are both between 0 and 1, and there are measurements of these quantities for many neural systems. Usrey and colleagues, for instance, found efficacies ranging 0.6$\%$ to 36$\%$ for retinogeniculate synapses \citep{Usrey_Reppas_Reid_1999} and 0.1$\%$ to $\sim 30\%$ for thalamocortical synapses \citep{Usrey_Alonso_Reid_2000} in cats. Swadlow and colleagues have found similar efficacies in rabbit thalamocortical projections \citep{Zhuang_Stoelzel_Bereshpolova_Huff_Hei_Alonso_Swadlow_2013} with some powerful, divergent connections having efficacies up to 20$\%$ \citep{Swadlow_Gusev_2002}. Putative synapses in the early auditory pathway have strong connections with both efficacies and contributions between 1-10$\%$ \citep{Miller_Escabí_Read_Schreiner_2001,Young_Sachs_2008}, and calyceal synapses in the mammalian auditory midbrain can even have efficacies >50$\%$ \citep{Guinan_Li_1990,Keine_Rubsamen_Englitz_2016}. With large-scale recordings, \cite{English_McKenzie_Evans_Kim_Yoon_Buzsáki_2017} found efficacies from 0.1$\%$ to $\sim 20\%$ in the hippocampus, while \cite{Sibille_Gehr_Benichov_Balasubramanian_Teh_Lupashina_Vallentin_Kremkow_2022} found median efficacies  of $\sim 3\%$ for retinal projections to superior colliculus. These examples provide only a snapshot of a few neural systems, but they illustrate a fundamental point: single presynaptic inputs can modify postsynaptic spiking probability by several percentage points in typical cases and up to 10’s of percentage points in extreme cases. Synaptic efficacies are, thus, often large enough that individual synapses can be studied using spike observations.

Although our goal is to estimate causal synaptic effects \citep{Hernán_2018}, there are limits to what we can conclude about synapses from spike observations alone. It is important to note that we generally cannot infer the presence of a connection with complete confidence from only the observed cross-correlation \citep{Brody_1999,Gerstein_Bedenbaugh_Aertsen_1989,Stevenson_Rebesco_Miller_Körding_2008} or rule out the possibility of a connection based on the absence of a correlation. Interpreting observed cross-correlation (or other measure of “functional connectivity”) as a direct estimate of causal impact would be a mistake \citep{Mehler_Kording_2020}. There are experimental methods for empirically verifying the presence of synapses, such as antidromic stimulation, juxtacellular/micro- stimulation, optogenetic stimulation, or by applying synaptic blockers \citep{English_McKenzie_Evans_Kim_Yoon_Buzsáki_2017,Sherman_Usrey_2021,Sibille_Gehr_Benichov_Balasubramanian_Teh_Lupashina_Vallentin_Kremkow_2022,Swadlow_Lukatela_1996}. Without such manipulations or without being able to otherwise control for possible confounds \citep{Lepperød_Stöber_Hafting_Fyhn_Kording_2022}, efficacy estimates are unverified, indirect, and potentially biased estimates of causal synaptic effects. However, even without verification, it is worth considering that, for some recorded neurons, their spike timing is almost certainly impacted by the presence of synapses with other recorded neurons. Models of synaptic effects may be the most parsimonious explanation for rapid, transient changes in postsynaptic spike probability following presynaptic spikes. Indirect or “circumstantial”  evidence from other sources, such as spike waveforms and anatomical location, can also strengthen (or weaken) our belief that a given pair of neurons is synaptically connected. In general, strong synaptic connections will impact the timing, shape, stability, and dynamics of pairwise spike statistics when they are present. Accurate models of neural circuits should be able to explain these effects, and fitting models based on synaptic connections may be fruitful, as long as we keep in mind that there are errors and biases in synapse detection.

Acknowledging the limitations of modeling putative synapses from spikes, there are many interesting scientific questions that we can at least begin to answer using observation alone. How does synaptic strength change with the precise timing of presynaptic activity? Are the strengths of synapses stable in vivo? How do synaptic strengths vary with behavior or stimuli or during learning? How might presynaptic spikes contribute to a postsynaptic neuron’s tuning properties? How does spike transmission through specific microcircuits change under different brain states? And many others. Although we may not be able to completely address these questions without confirmation from intracellular or other measurements, the efficacies of putative synapses provide initial (potentially biased) estimates of causal synaptic effects, and models based on synaptic effects provide potential explanations that can be tested and compared against alternative models. Previous studies have aimed to describe the detailed shape of spike cross-correlations at putative synapses \citep{Fetz_Gustafsson_1983,Herrmann_Gerstner_2001, Herrmann_Gerstner_2002,Poliakov_Powers_Sawczuk_Binder_1996}, the dynamics of membrane potential integration \citep{Carandini_Horton_Sincich_2007}, short-term plasticity \citep{English_McKenzie_Evans_Kim_Yoon_Buzsáki_2017,Ghanbari_Malyshev_Volgushev_Stevenson_2017,Ghanbari_Ren_Keine_Stoelzel_Englitz_Swadlow_Stevenson_2020}, behavioral and brain state dependencies \citep{Csicsvari_Hirase_Czurko_Buzsáki_1998,Fujisawa_Amarasingham_Harrison_Buzsáki_2008,Senzai_Fernandez-Ruiz_Buzsáki_2019,Stoelzel_Bereshpolova_Swadlow_2009}, and receptive field/tuning curve construction \citep{Alonso_Swadlow_2005,Maurer_Cowen_Burke_Barnes_McNaughton_2006,McKenzie_Huszár_English_Kim_Christensen_Yoon_Buzsáki_2021,Miller_Escabí_Read_Schreiner_2001,Reid_2012}. In each of these studies, models of synaptic effects generate quantitative predictions and act as key tools to interpret the observed correlations.

Here we provide an overview of methods and logic for modeling synapses from large-scale spike train observations. Although the specific models and concepts here have been previously described, the aim of this paper is to illustrate how these varied approaches can be synthesized into a coherent data analysis framework. The results that follow focus on new case-studies from experimental data that directly demonstrate key ideas. Interspersed with these results, however, we discuss conceptual challenges and highlight previous studies that introduced or expanded on these challenges studying synapses from spikes. In the first section, we consider the basic theoretical problem of when synapses can be detected from spike observations alone. This detection problem is often framed as a hypothesis test, and a power analysis provides useful “rules of thumb” for distinguishing when synapses can be detected and when they cannot. Second, we discuss a conceptual framework for interpreting synaptic effects beyond the hypothesis test using circumstantial evidence. Third, after discussing circumstantial evidence broadly, we use case studies to directly illustrate the impact of specific circumstantial evidence, such as anatomy and spike waveforms. Fourth, we illustrate some of the ways that existing model-based approaches can account for detailed patterns of pairwise spike statistics and describe synaptic plasticity. Fifth, we discuss concerns about confounding due to common input and polysynaptic effects. Lastly, we illustrate how putative synapses detected in large-scale spike recordings are reaching a scale where they might allow observing and characterizing putative microcircuits.

\begin{figure}[bt]
\centering
\includegraphics[width=\textwidth]{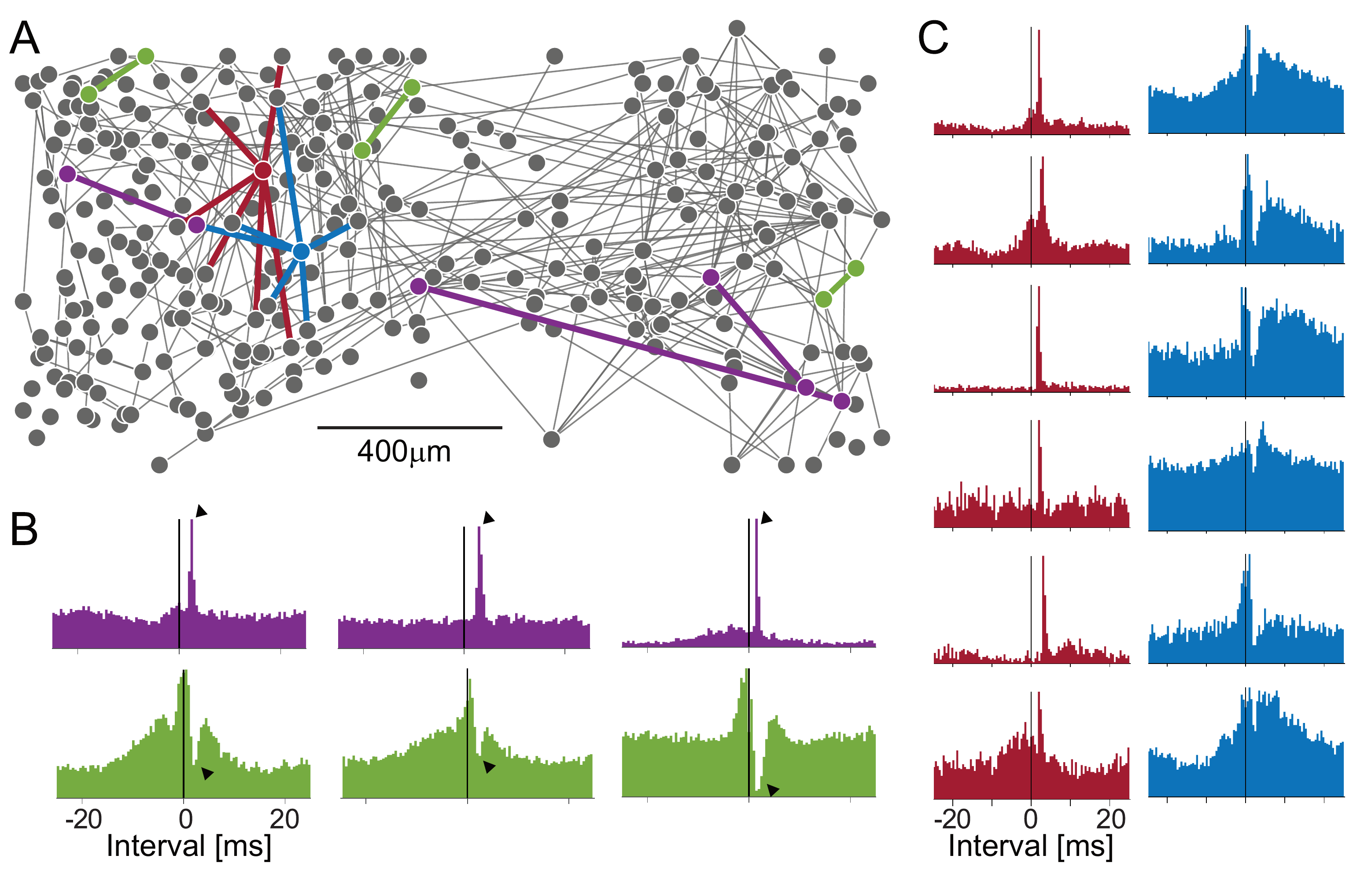}
\caption{Example putative synaptic connections detected from a multi-electrode array spike recording of a slice culture of mouse somatosensory cortex (CRCNS ssc3-23). A) Estimated network structure with single units aligned to their spatial locations on the multi-electrode array. B) Example cross-correlations from strong putative excitatory (purple) and inhibitory (green) monosynaptic connections. Triangles denote the peak or trough of the putative synaptic effect. C) Multiple putative synapses from the same presynaptic neuron are often either all excitatory or all inhibitory, consistent with Dale’s Law. Colors in (B) and (C) correspond to the edges shown in (A).}
\end{figure}

\section{Results}

We start by considering the primary statistical evidence for the presence of a synaptic connection: the cross-correlogram. Here the spiking of a putative postsynaptic neuron is compared to the spikes of a putative presynaptic neuron. Given the binned spiking of the two neurons $n_{pre}$ and $n_{post}$ over time, the cross-correlogram is

\begin{equation*}
c(\tau)=\sum_t {n_{pre} (t) n_{post} (t+\tau)}
\end{equation*}

Increases or decreases in $c(\tau)$ at a delay $\tau$ between 0ms and (approximately) 5-10ms may reflect synaptic effects when a synapse is present. Detecting a short-latency, transient deviation in $c(\tau)$ relative to an expected baseline provides the basis for many synapse detection methods \citep{Amarasingham_Harrison_Hatsopoulos_Geman_2012,Moore_Segundo_Perkel_Levitan_1970}. Typically, the observed count within a specific range of $\tau$ is compared to a null hypothesis that describes what is expected to happen in the absence of a synapse between the neurons. If the observed count is unlikely under the null hypothesis, this provides some evidence that there may be synaptic effect.

To give some initial, concrete examples of what putative synaptic effects look like and a sense of how these effects are detected we introduce a case-study (Fig 1). Here, we analyze spike data from an in vitro multi-electrode array recording of mouse somatosensory cortex. There are $M=310$ neurons with $M(M-1)$ possible (directed) connections, $\sim$100K in this recording. After testing all possible pairs, we can construct a putative synaptic network for the observed sample of units (Fig 1A). A small subset of pairs have a short-latency transient increase in postsynaptic firing following presynaptic spikes (Fig 1B, top), while others show a short-latency transient decrease (Fig 1B, bottom). The latency, timescale, and efficacy of these patterns are often consistent with expected synaptic effects.

\subsection{Power analysis for the detection of synapses from spikes}

Since an individual synapse represents only one of, potentially, many inputs to a postsynaptic neuron, it is reasonable to ask: when can the impact of a single input be reliably detected and measured from spikes alone? Detecting a synaptic connection from a noisy correlogram can be framed as a null hypothesis significance test, and, as with other hypothesis tests, power analysis can be used to determine under what circumstances real effects are expected to result in statistically significant outcomes. Given an effect size, a sample size, and a confidence level for the hypothesis test $\alpha$, we can calculate the probability of having a false negative (Type II error) $\beta$. Or, conversely, given a sample size, confidence level, and desired power, we can calculate the effect size necessary to achieve that power. 

Let’s consider a simplified situation where we aim to detect a synaptic connection with efficacy $e$ given $N$ presynaptic spikes and known probability p for the postsynaptic neuron firing by chance within a fixed detection window following each presynaptic spike. We wish to decide between two hypotheses: $H_0:e=0$ versus $H_A:e\neq 0$. Assuming that the observations are independent, the null distribution for the number of postsynaptic spikes in the detection window is given by $y \sim Binomial(N,p)$ and the alternative distribution is $y \sim Binomial(N,p+e)$ (Fig 2A). Testing in this situation is analogous to a one-sample test of a proportion, and the power $(1-\beta)$ can be calculated exactly. Fig 2B illustrates how the power varies with the effect size ($e$) and sample size ($N$). For experimentally plausible values ($\sim$1000 presynaptic spikes), the power is often high enough that even synapses with relatively low efficacy (5$\%$) could be reliably detected. 

Similarly, we can identify the minimal efficacy necessary to achieve a specific power (Fig 2C). With some simplifying approximations, an approximate threshold of detectable effects is $e\approx 1.4/\sqrt N$ (for the commonly used values of confidence level $\alpha=0.05$ and power $0.8$, see Methods). That is, to have an $\ge80\%$ chance of detecting a synaptic connection after observing 100, 1000, or 10000 presynaptic spikes, the efficacy would need to at least 14$\%$, 4$\%$, or 1$\%$, respectively. This approximate power analysis has important caveats; however, it provides a rule of thumb linking the number of recorded spikes to the problem of synapse detection.

Several more complex hypothesis test-based methods for detecting putative synapses \citep{Amarasingham_Harrison_Hatsopoulos_Geman_2012,Fujisawa_Amarasingham_Harrison_Buzsáki_2008,Liew_Pala_Whitmire_Stoy_Forest_Stanley_2021,Spivak_Levi_Sloin_Someck_Stark_2022}, as well as several model-based methods for detecting synapses have been recently developed \citep{Endo_Kobayashi_Bartolo_Averbeck_Sugase-Miyamoto_Hayashi_Kawano_Richmond_Shinomoto_2021,Kobayashi_Kurita_Kurth_Kitano_Mizuseki_Diesmann_Richmond_Shinomoto_2019,Ren_Ito_Hafizi_Beggs_Stevenson_2020, Zaytsev_Morrison_Deger_2015}. These methods are more involved than the simple hypothesis test presented above. The jitter method \citep{Amarasingham_Harrison_Hatsopoulos_Geman_2012}, for instance, relies on an empirical null distribution generated by adding timing noise to the spikes, and GLMCC \citep{Kobayashi_Kurita_Kurth_Kitano_Mizuseki_Diesmann_Richmond_Shinomoto_2019} detects pairs based on a model goodness-of-fit. Despite their differences, the power, false positive, and false negative rates of these models can be estimated when the ground-truth connectivity is known, such as in simulations with controlled synaptic currents or potentials \citep{Kobayashi_Kitano_2013,Kobayashi_Kurita_Kurth_Kitano_Mizuseki_Diesmann_Richmond_Shinomoto_2019,Moore_Segundo_Perkel_Levitan_1970,Ren_Ito_Hafizi_Beggs_Stevenson_2020,Volgushev_Ilin_Stevenson_2015}. These methods give slightly different putative synaptic networks and their results depend on the specified confidence level, but there is generally agreement across methods for the pairs where the strongest effects are present. These studies, along with experimental results from identified synapses \citep{English_McKenzie_Evans_Kim_Yoon_Buzsáki_2017,Sibille_Gehr_Benichov_Balasubramanian_Teh_Lupashina_Vallentin_Kremkow_2022}, suggest that synapses can indeed be detected in vivo from only spikes. However, weak synapses are less likely to be detected than strong synapses, synapses between neurons with low firing rates are less likely to be detected than those between neurons with high firing rates, and excitatory synapses may not necessarily be detected at the same rate as inhibitory synapses \citep{Aertsen_Gerstein_1985,Kobayashi_Kurita_Kurth_Kitano_Mizuseki_Diesmann_Richmond_Shinomoto_2019,Palm_Aertsen_Gerstein_1988,Volgushev_Ilin_Stevenson_2015}.

\begin{figure}[bt]
\centering
\includegraphics[width=\textwidth]{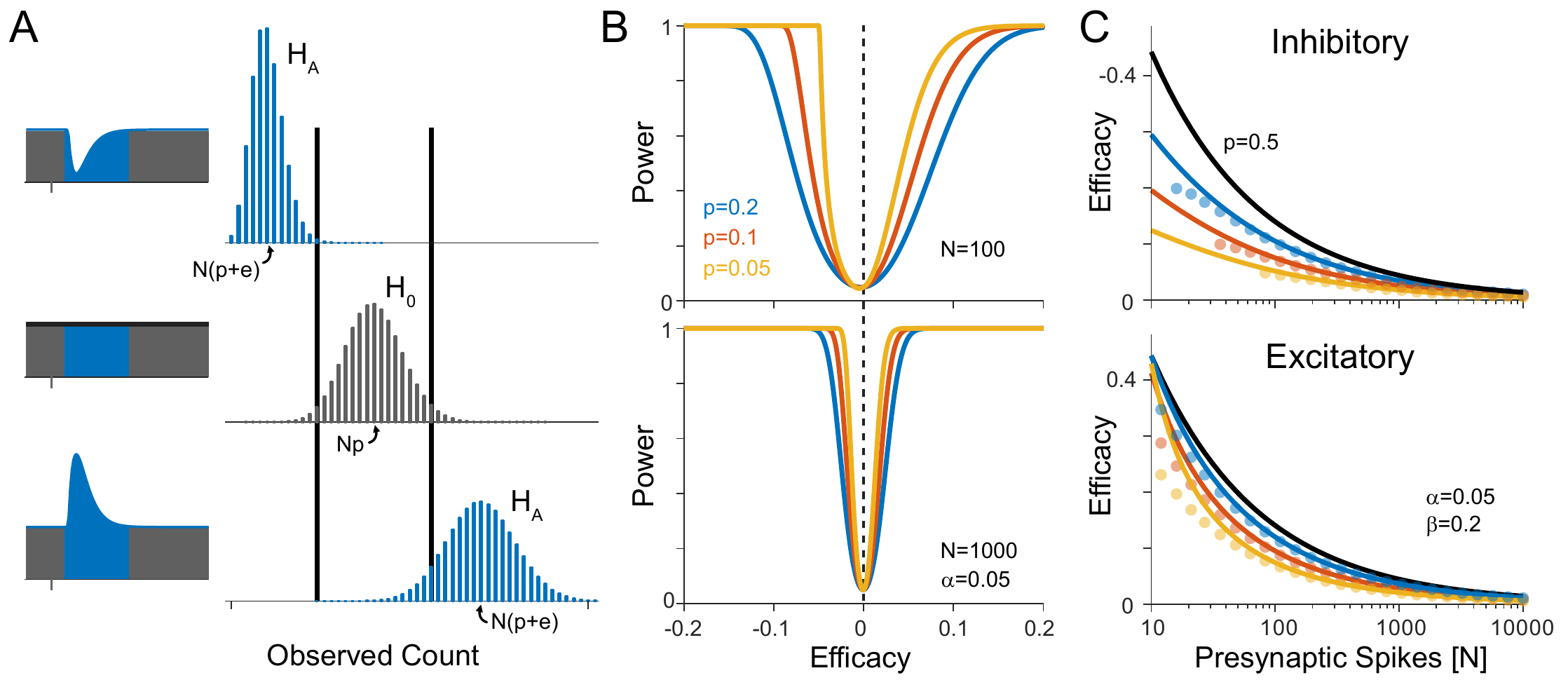}
\caption{Power analysis for detecting synaptic connections with known baseline and unknown efficacy. A) Here we consider a simplified hypothesis test between a null hypothesis and an alternative where a synapse increases (bottom) or decreases (top) the probability of postsynaptic spiking by a specific efficacy. We assume that the window of interest is already known, and that we run our test on a single observed count of the postsynaptic spikes. Vertical lines denote decision boundaries for a two-tailed hypothesis test. B) Using a normal approximation to the Binomial distribution we can calculate the power associated with a specific efficacy. Here curves show the power for multiple baseline firing probabilities $p$ at $N=100$ (top) and $N=1000$ (bottom) presynaptic spikes. C) Using this approximation, we can also find the efficacy necessary to achieve a desired power. Here we show efficacies need to achieve a power of 80$\%$, with inhibitory (top) and excitatory (bottom) synapses. Dots denote calculations from the normal approximation to the binomial. Curves denote the first-order Taylor approximation. In all cases shown here we assume a confidence level $\alpha$=0.05.}
\end{figure}

\subsection{Evaluating circumstantial evidence for synaptic connections}
We now have a statistical argument that many synapses can, in principle, be detected from spikes when they are present, and a statistically significant transient in the correlogram provides some direct evidence of a potential causal effect. However, we should be cautious in concluding that a synapse is present from observations of spikes alone. In the absence of electrophysiological, optogenetic, or pharmacological manipulations to verify that these are synaptic effects, we cannot be sure that short-latency, transient changes in spiking probability are “caused” by a synapse. Here we briefly review the indirect, but multiple, lines of evidence that can, nonetheless, strengthen or weaken our belief that pairs of simultaneously recorded neurons may be monosynaptically connected.

First, we consider the detailed shape of the cross-correlogram. Although many studies of functional connectivity do not explicitly distinguish between synaptic effects and other possible sources of fast correlated spiking \citep{Chen_Putrino_Ghosh_Barbieri_Brown_2011,Ito_Yeh_Hiolski_Rydygier_Gunning_Hottowy_Timme_Litke_Beggs_2014,Stevenson_Rebesco_Miller_Körding_2008}, when studying putative synapses directly, researchers often aim to measure the latency and time scale of the synaptic effect \citep{Swadlow_Gusev_2001,Usrey_Reppas_Reid_1999}. Due to the conduction velocity and synaptic delay, synaptic effects, when present, will have a short, but non-zero latency. Then, due to the receptor dynamics (AMPA, NMDA, GABAA, and/or GABAB), the timescale of the synaptic effect will be limited to a short duration window following the presynaptic spikes. Both experimental \citep{Fetz_Gustafsson_1983,Poliakov_Powers_Sawczuk_Binder_1996} and theoretical \citep{Herrmann_Gerstner_2001, Herrmann_Gerstner_2002} studies have examined the transformation between synaptic potentials/currents and postsynaptic spike probability, and these results provide constraints on what is possible with synaptic transmission. A transient increase in the correlogram with a latency <10 ms and a duration of <3ms, such as those in Fig 1B, is consistent with synaptic transmission. However, an increase with a latency of 20-50ms and a duration of >20ms is somewhat inconsistent with our expectations for monosynaptic transmission via ionotropic receptors for small mammals.

A second source of circumstantial evidence comes from anatomy and histology. The probability of two neurons being monosynaptically connected depends, crucially, on their locations. A neuron recorded in the retina is not likely to be directly connected to a neuron recorded in the spinal cord, and structural connectivity, in general, constrains conclusions about putative synapses. In some cases, neurons are targeted for their receptive field alignment \citep{Usrey_Alonso_Reid_2000,Zhuang_Stoelzel_Bereshpolova_Huff_Hei_Alonso_Swadlow_2013}. Using information about receptive field overlap, which brain areas are connected to which \citep{Harris_Mihalas_Hirokawa_Whitesell_Choi_Bernard_Bohn_Caldejon_Casal_Cho_et_al._2019,Köbbert_Apps_Bechmann_Lanciego_Mey_Thanos_2000}, or simply the distance between the recorded neurons can additionally constrain the inference of synaptic connections. Even within brain areas, intracellular measures often find that connection probability decreases with distance and latency grows with distance \citep{Seeman_et_al._2018,Song_Sjostrom_Reigl_Nelson_Chklovskii_2005}. Putative synapses detected from extracellular recordings often mirror these trends \citep{English_McKenzie_Evans_Kim_Yoon_Buzsáki_2017}, and the trends can be used to constrain our detection \citep{Ren_Ito_Hafizi_Beggs_Stevenson_2020}.

Extracellular spike waveforms offer another important piece of circumstantial evidence. Since cell types differ in their waveform characteristics \citep{Lee_Balasubramanian_Tsolias_Anakwe_Medalla_Shenoy_Chandrasekaran_2021,Trainito_von_Nicolai_Miller_Siegel_2019}, we may have clear expectations about the direction (increasing or decreasing spike probability) of synaptic effects at individual putative connections even before we examine the cross-correlogram. Putative presynaptic units with broad spike waveforms, often pyramidal neurons, tend to have excitatory putative synaptic effects, while units with narrow spike waveforms, often fast-spiking interneurons, tend to have inhibitory effects  \citep{Barthó_Hirase_Monconduit_Zugaro_Harris_Buzsáki_2004}. More detailed approaches to cell type classification using more complex waveform features or the full, multi-channel spatio-temporal waveform pattern \citep{Delgado_Ruz_Schultz_2014,Li_Gauthier_Schiff_Sher_Ahn_Field_Greschner_Callaway_Litke_Chichilnisky_2015,Shein-Idelson_Pammer_Hemberger_Laurent_2017} are also being actively developed.

Next, we consider the patterns of putative synapses beyond a single pair. In a large-scale multi-electrode recording, we may have many putative synapses that are detected, and the pattern across these many connections provides circumstantial evidence for or against the individual pairs of neurons being connected. Dale’s Law, that neurons are either excitatory or inhibitory \citep{Strata_Harvey_1999}, provides an important example. A putative presynaptic neuron, if it is excitatory, is expected to make excitatory connections with all its postsynaptic targets, while an inhibitory presynaptic neuron should make inhibitory connections. Despite growing evidence that co-transmission of neurotransmitters can occur \citep{Tritsch_Granger_Sabatini_2016}, we often find that Dale’s Law does hold (approximately) when examining putative synapses (Fig 1C), and other details about connection probabilities in local circuits \citep{Haber_Wanner_Friedrich_Schneidman_2023,Song_Sjostrom_Reigl_Nelson_Chklovskii_2005} could also be used.

Lastly, we consider examining spike statistics in more detail. The overall cross-correlogram summarizes the pairwise spike statistics across the whole of an experimental recording. In most cases, using all spikes that are available will give us the greatest statistical power for detecting putative connections. However, the overall cross-correlogram is only one limited summary of the dependency between pre- and postsynaptic spiking. Analyzing the data during specific time periods, behaviors, brain states, or specific patterns of pre-synaptic activity can provide a richer understanding of the pre-post spike dependencies. On the one hand, we expect that when a synapse is present, certain features of the spike statistics should be stable no matter how we look. The latency and duration of the synaptic effect, for instance, are expected to be relatively stable across time and should not change substantially across behaviors or brain states (although, see \cite{Bakkum_Chao_Potter_2008}). On the other hand, we expect that other features of the spike statistics should systematically vary if there is a synapse. The postsynaptic spike probability may show evidence of nonlinearities due to membrane potential integration, such as different probabilities of spiking following a burst vs an isolated presynaptic spike, and, due to short-term synaptic plasticity, the efficacy of a true synaptic connection is expected to increase or decrease rapidly, depending on the pattern of presynaptic activity. These phenomena have both been observed with putative synapses \citep{Carandini_Horton_Sincich_2007,English_McKenzie_Evans_Kim_Yoon_Buzsáki_2017,Swadlow_Gusev_2001}. Importantly, however, both the stability and variation of the correlogram reflect predictable consequences of synaptic biophysics. When they are present, they may provide additional support for our belief that a putative synapse is a genuine synapse.

\begin{figure}[bt]
\centering
\includegraphics[width=\textwidth]{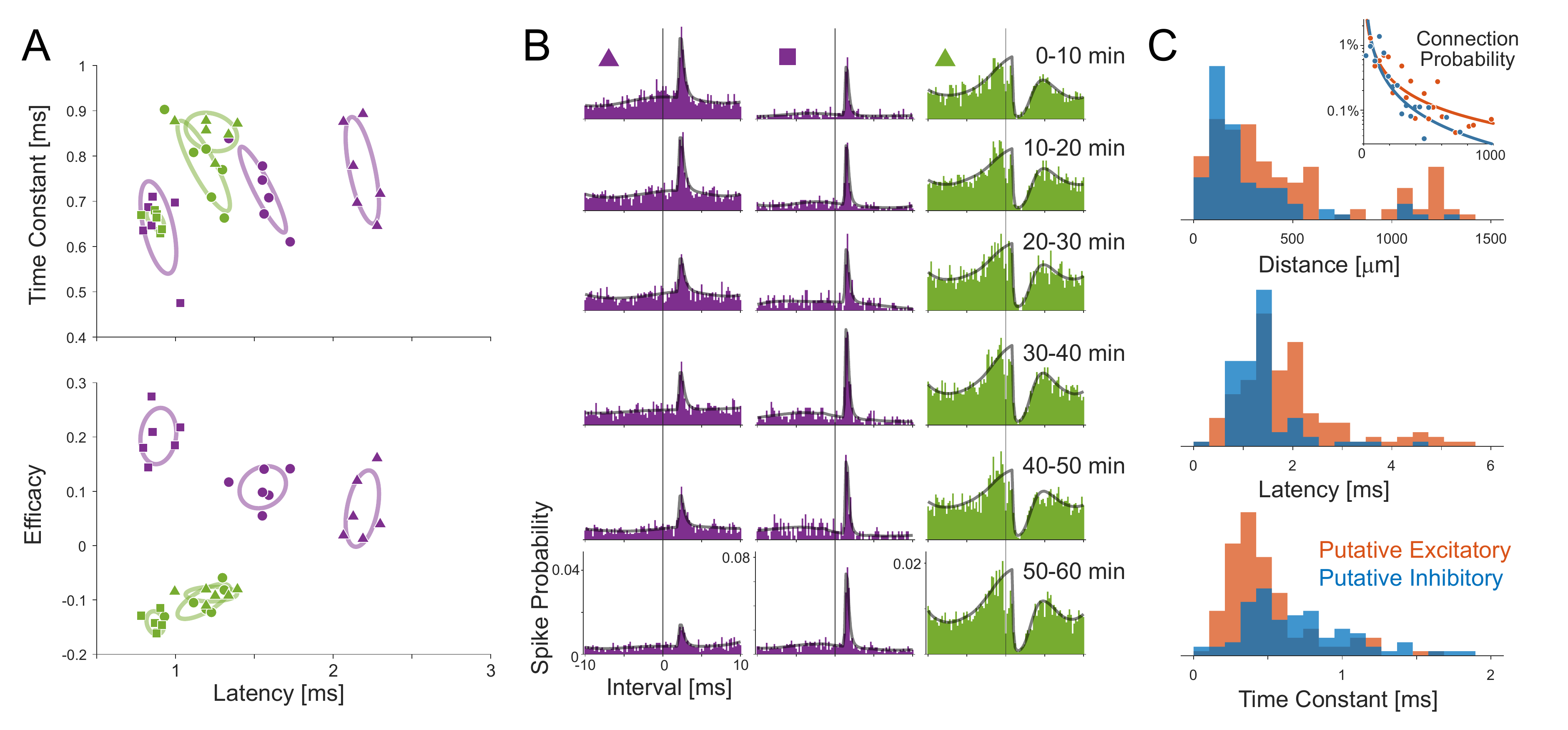}
\caption{Parameters and stability of putative synaptic effects for CRCNS SSC-3 dataset 23. A) Latency, time constant, and efficacy estimated for 10-minute segments of the 1-hour recording. Symbols denote the 6 putative synapses shown in Fig 1B, with putative excitatory connections in purple and putative inhibitory connections in green. Ellipses denote estimated mean and covariance. B) Correlograms for the 10-minute segments of data, along with a model fit (gray) that assumes fixed latency and time constant. C) Across all putative synapses, there are consistent differences in the distances between pairs, latencies, and time constants for excitatory and inhibitory effects. Connection probability (inset) for both effects decreases with distance between the pairs. Lines denote exponential fit within a Binomial regression of observed counts to possible pairs at a given distance.}
\end{figure}
Descriptions of the pre-post spike statistics beyond the overall correlogram also allow models of synaptic effects to be explicitly tested. After fitting a synaptic model to one subset of the data, the model predictions can be evaluated on other subsets. In previous work we found that detailed models of short-term synaptic dynamics can predict responses to triplets of presynaptic spikes \citep{Ghanbari_Ren_Keine_Stoelzel_Englitz_Swadlow_Stevenson_2020}, slow rate-dependent fluctuations in efficacy \citep{Ren_Wei_Ghanbari_Stevenson_2022}, and differences across stimuli \citep{Ghanbari_Ren_Keine_Stoelzel_Englitz_Swadlow_Stevenson_2020}. Multiple models can be compared directly, and individual models of synaptic effects are falsifiable, in the sense that their predictions may be wrong.

\subsection{Case-studies evaluating putative synapses in large-scale spike recordings}

To illustrate how multiple lines of circumstantial evidence can fit together we further examine results from two experimental datasets: 1) the 1hr recording from organotypic slice cultures of mouse somatosensory cortex introduced above (Collaborative Research in Computational Neuroscience SSC-3 dataset 23) and 2) a 2.7hr multi-region recording from awake mouse using multiple Neuropixels arrays (Allen Institute – Visual Coding Neuropixels electrophysiology session 715093703). SSC3-23 contains M=310 sorted, single units (automatic clustering based on PCA followed by manual refinement) with median firing rate 1.8 Hz [0.8, 3.7] (quartiles). ABI-715093703 contains 745 sorted, single units (SNR>2, Kilosort 2) with median firing rate 6.7 Hz [3.1, 10.7].

We exclude single-units with fewer than 1000 spikes (0.28 Hz, 6.7$\%$ in SSC3-23, 0.10 Hz, 0.3$\%$ in ABI-715093703), since even moderate putative connections with efficacies $\sim4\%$ would be difficult to detect with so few spikes. This leaves $M'=289$ units for SSC3-23 and $M'=743$ units for ABI-715093703. Based on the approximate power analysis described above, we then assess limits on detecting putative synapses with 80$\%$ power. Assuming $\alpha=0.05$, and $\beta=0.2$ and a 10ms synaptic window following presynaptic spikes to estimate p, the median efficacy detection limit across all pairs is ±0.3$\%$ (ABI-715093703) and ±0.5$\%$ (SSC3-23). That is, in these relatively long recordings we expect efficacies less than $\sim 1\%$ to be detectable for typical pairs of units.

Here, rather than comparing or optimizing detection, we want to highlight how circumstantial evidence can shape interpretation of cross-correlation observations. We first detect candidate pairs using a fast approximation to the jitter method (jitter timescale of 2ms) and correct for multiple comparisons using Benjamini-Hochberg step-up procedure with False Discovery Rate of $\alpha=0.1$ (see Methods). In SSC3-23 this step identifies 347 pairs out of 83K possible pairs (0.4$\%$, Fig 1A) as having statistically significant fast correlations. This approach detects deviations from expected slow correlations but does not identify effects that are consistent or inconsistent with putative synapses. We thus fit the correlograms for statistically significant pairs with a model that separates the correlogram into a slow, baseline component and a fast, synaptic effect. Candidate correlograms are often well-fit by such a model (median pseudo-$R^2$ is 0.74), and the estimated parameters are often consistent with synaptic timescales (Fig 3). Since some candidate correlograms are not well fit and/or not consistent with synaptic effects, we next restrict our analyses to more specific parameter ranges. Here we focus on pairs with pseudo-$R^2>0.5$, time constant<2ms, and latency between 0.1-6ms with 213 pairs meeting these criteria (61$\%$ of candidates, 0.3$\%$ of all possible pairs).

\begin{figure}
\centering
\includegraphics[width=\textwidth]{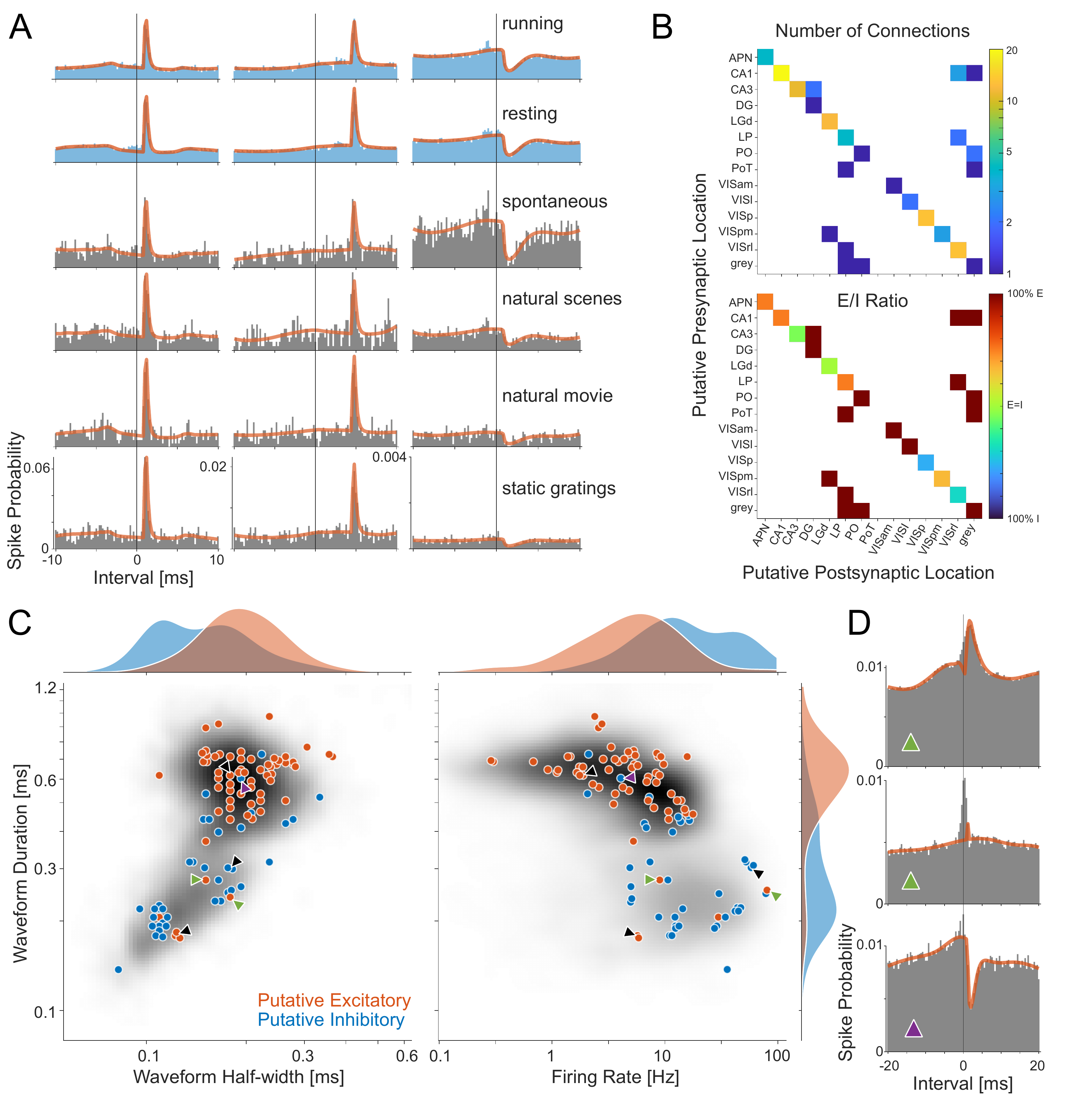}
\caption{Stability, anatomy, waveforms of putative synaptic effects of ABI-715093703. A) Correlograms for three example putative synapses during different behavior and visual stimuli. Blue and gray bars denote observed count data, red curves denote a model fit that assumes fixed latency and time constant (but variable baseline and efficacy) across stimuli. Left column is a pair of units in CA1, middle column is a putative presynaptic unit in CA3 and postsynaptic unit in DG, right column is a pair in VISp. B) Connectivity matrix for the putative connections detected from this dataset (top), and ratio of putative excitatory to inhibitory pairs for each projection (bottom). C) Waveform and firing rate statistics for the putative presynaptic neurons. Putative excitatory connections tend to be from presynaptic neurons with wider waveforms and to have lower firing rates. Grayscale background denotes the histogram of all units across all ABI recordings. Density plots along edges denote kernel estimates for ABI-715093703 separated by putative excitatory connections (red) and putative inhibitory connections (blue). Black triangles denote the example pairs in (A). Green and purple triangles denote the example pairs in (D). D) Correlograms for example pairs with inconsistent waveform and correlogram information (pairs are in CA1 top, VISpm middle, and VISp bottom).}
\end{figure}

Having latencies and time constants consistent with synaptic effects may be sufficient to label these pairs of neurons “putative synaptic connections”, but additional evidence can help us answer “how putative are we talking?” and “is this pair worth studying more?”. The stability of the putative synaptic effect is one piece of circumstantial evidence that can inform both questions. Here, to assess the stability of the effects, we split the data into 10-minute windows and estimate synaptic parameters for each segment of the data. We find that, in many examples, latency and duration both tend to be reliable over time (Fig 3A) with estimates varying by <1ms across segments. Efficacy, as well as the baseline postsynaptic firing rate, can vary more substantially (on the order of ±50$\%$), but typically the sign of the putative synaptic effect does not change (Fig 3B). Variation in the correlograms over time can be well described by a model that assumes that the latency and time constant of the fast effect are fixed, while the efficacy and baseline vary.

In deciding whether a set of putative pairs is worth studying further we can check whether features across all pairs align with our expectations. In SSC3-23, the overall efficacies for putative excitatory connections ($n=128$) and putative inhibitory connections ($n=85$) are comparable (median of 4$\%$ [2$\%$, 7$\%$] for excitatory, -4$\%$ [-2$\%$, -5$\%$] for inhibitory, with intervals denoting quartiles). However putative excitatory connections tend to occur between neurons whose distance on the electrode array is further apart (median 308 $\mu m$ [168, 581] quartiles for excitatory and 168 $\mu m$ [119, 338] for inhibitory, Wilcoxon rank-sum $p<10^{-4}$). Time constants for putative excitatory connections tend to be shorter (median 0.4 ms [0.3, 0.6] quartiles for excitatory and 0.6 ms [0.5, 0.9] for inhibitory, rank-sum $p<10^{-6}$), and latency tends to be longer (median 1.7 ms [0.2, 2.2] for excitatory and 1.3 ms [1.1, 1.5] for inhibitory, rank-sum $p<10^{-4}$). We also find that connection probability decays with distance (Fig 3C, inset), and the length scale for putative excitatory connections is somewhat longer (Binomial regression for observed counts to possible pairs at a given distance, OR=0.50 [0.43, 0.57] per doubling for excitatory connections 0.4 [0.35, 0.47] per doubling for inhibitory connections, intervals denote 95$\%$ CI). All these observations are consistent with intracellular observations of intracortical synaptic effects.

To illustrate some additional ways that circumstantial evidence can inform the interpretation of individual correlograms, we switch our focus to ABI-715093703 – a large-scale, multi-region recording with six Neuropixels arrays from an awake, head-fixed mouse viewing a varying visual stimulus. Here the detection step identifies 201 pairs out of 551K possible pairs (0.04$\%$) as having statistically significant fast correlations. We again fit correlograms for candidate pairs with a model of putative synaptic effects (median pseudo-R2 is 0.79) and restrict analyses to more specific parameter ranges (criteria on latency, time constant, and pseudo-$R^2$ as above, 51$\%$ of candidates passing, 0.02$\%$ of all possible pairs). Here we find that the efficacies for putative excitatory connections ($n=63$) are somewhat stronger than those for putative inhibitory connections (n=39) (median of 4$\%$ [2$\%$, 10$\%$] for excitatory, -1$\%$ [-0.6$\%$, -1.8$\%$] for inhibitory). Time constants for putative excitatory connections tend to be shorter (median 0.58 ms [0.37, 0.78] for excitatory and 0.74 ms [0.43, 1.08] for inhibitory, rank-sum $p=0.06$), and latency tends to be longer (median 1.0 ms [0.8-1.5] for excitatory and 0.9 ms [0.8-1.1] for inhibitory, rank-sum $p=0.04$).

As with SSC3-23, we can ask whether putative synaptic effects are stable. In this case, since these recordings are from awake, behaving mice, we assess whether they are stable across stimuli and behavior. We again find that, in many examples, latency, duration, and sign (excitatory or inhibitory) all tend to be reliable while efficacy and baseline postsynaptic firing rate vary (Fig 4A). Since these recordings sample multiple brain regions and multiple arrays, we also quantify differences in putative synaptic connectivity across anatomy (Fig 4B). Most putative synapses (88$\%$) come from neurons recorded from the same Neuropixels probe and brain area (84$\%$). The average distance between putative excitatory pairs on the same probe is slightly longer (84 ± 59 $\mu m$, S.D.) than the distance between putative inhibitory pairs (72 ± 48 $\mu m$), although not statistically significant (unpaired t-test $t(87.6)=1.1$, $p=0.29$). Of the inter-probe ($n=12$) and inter-area connections ($n=16$), all connections are putative excitatory connections. Inter-area connections could also be compared to anatomical/tracing data on connectivity \citep{Kuan_Li_Lau_Feng_Bernard_Sunkin_Zeng_Dang_Hawrylycz_Ng_2015} to verify if such connections genuinely occur or to strengthen/weaken our belief in these specific putative synapses

\begin{figure}
\centering
\includegraphics[width=\textwidth*9/10]{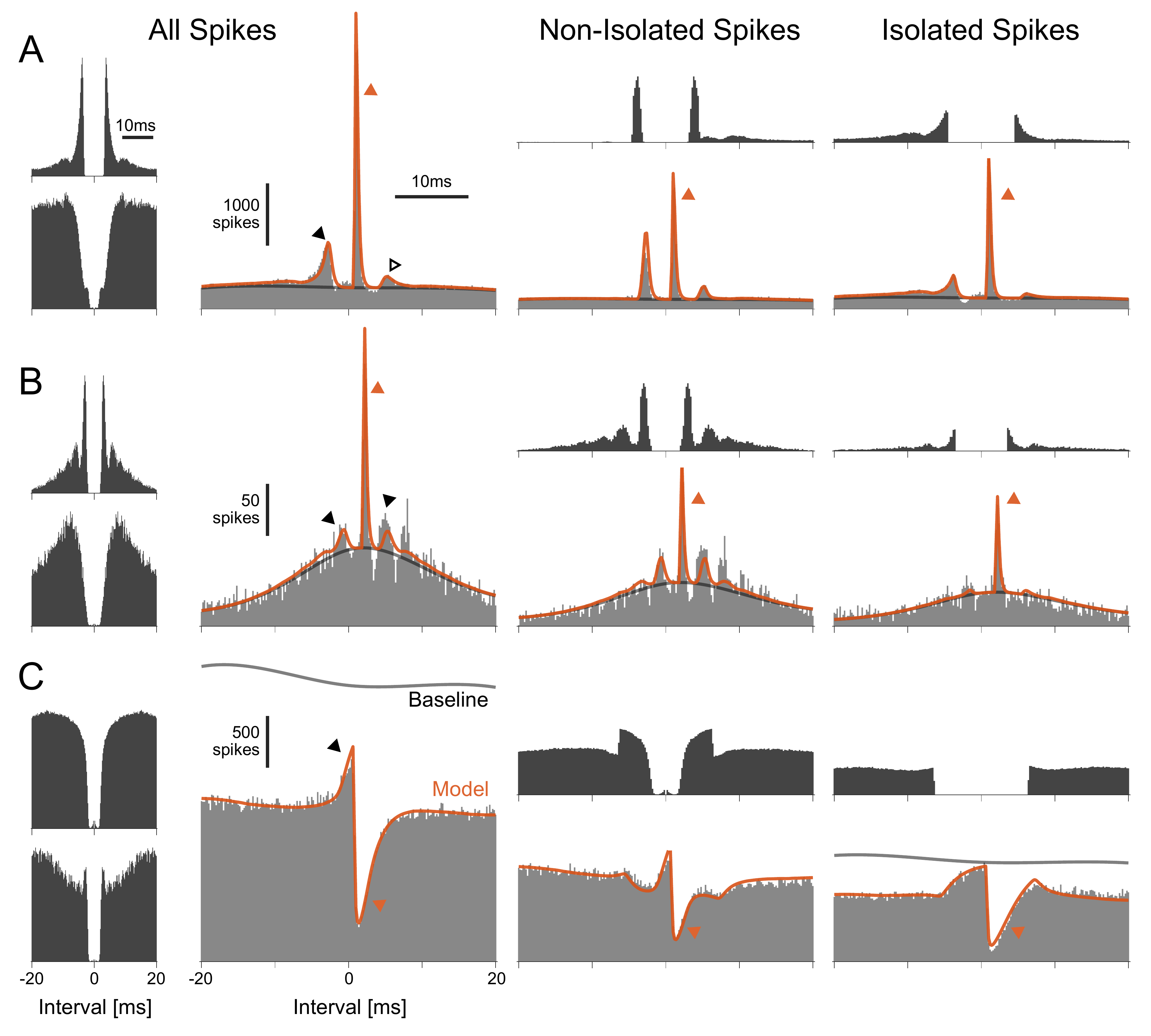}
\caption{Presynaptic dynamics can have a substantial influence on the observed cross-correlation. (A-C) show three of the most extreme examples taken from three recordings in the Allen Brain Observatory Neuropixels dataset. The auto-correlations of the presynaptic (top) and postsynaptic (bottom) neurons are shown at left, followed by the cross-correlation from all spikes (gray). We then compute cross-correlations separately for presynaptic spikes that were closely followed or preceded by another presynaptic spike (Non-Isolated) and presynaptic spikes that were Isolated from nearby presynaptic spikes by comparison (split by the median nearest spike time). (A) and (B) illustrate cases where bursting in the presynaptic neuron causes side-peaks in the cross-correlation (black triangles) in addition to the primary response (red triangle). These side-peaks are less pronounced for Isolated presynaptic spikes and can be asymmetric (filled vs open triangles in A). (C) Illustrates a case where the shape of the presynaptic autocorrelation causes a peak near an interval of 0, reflecting relief from inhibition (black triangle). In all cases the red curves denote the model fit, where a single synaptic effect is convolved with the presynaptic auto-correlation and acts in addition to a slowly fluctuating baseline (gray). Curves for Non-Isolated and Isolated spikes use the same synaptic model but with a case-specific shift in the baseline constant. Note that, in the auto-correlations, the peak at $\tau=0$ has been removed for clarity.}
\end{figure}

For this case study, we also want to demonstrate how spike waveform can play a role as circumstantial evidence. As many past studies have noted, spike waveforms are often clustered. Similar to previous work \citep{Barthó_Hirase_Monconduit_Zugaro_Harris_Buzsáki_2004}, here we find that spike waveforms form a pair of clusters and that the waveform-type (broad vs narrow) generally corresponds to the sign of the putative synaptic effect (Fig 4C). There are, however, some inconsistencies (Fig 4D). One reason these could occur may be if waveform does not uniquely identify cell type. Since narrow waveforms can occur when an electrode is close to an axon collateral \citep{Barry_2015}, and broad waveforms can occur for inhibitory neurons \citep{Moore_Wehr_2013}, particularly those that are “regularly spiking”, this is one possibility (e.g. Fig 4D, bottom). On the other hand, the synaptic effect may not always be accurately described despite high pseudo-$R^2$. With the model used here, the presence of fast, near-synchronous correlation, in particular, can lead to misestimation (Fig 4D, top and middle). Due to fact that the presynaptic neurons in these examples have narrow waveforms and the fact that the peak lacks  a clear synaptic latency, these examples are unlikely to be excitatory synapses. It is possible that these peaks are instead due to confounding common input (see below). Altogether, these case studies demonstrate how putative synaptic effects can be evaluated based not just on the overall correlogram, but also based on the stability of the correlogram, anatomical locations, spike waveforms, and the features of the population.

\subsection{Explanatory models, presynaptic spike auto-correlation, and short-term plasticity}
In evaluating putative synapses, we particularly want to highlight the value of explanatory models of synaptic effects. That is, models that 1) take presynaptic spike times as input and generate predictions about the postsynaptic spiking as output and 2) attempt to account for specific neurophysiological mechanisms. Within this class, modelers choose whether to predict detailed postsynaptic spike timing or to predict summary statistics, such as correlograms, and, additionally, choose whether to describe neuron dynamics using either membrane potentials or firing rates \citep{Bykowska_Gontier_Sax_Jia_Montero_Bird_Houghton_Pfister_Costa_2019}. Here we focus on rate-based models of the correlogram, but other descriptions may be necessary in more complex settings \citep{Ladenbauer_McKenzie_English_Hagens_Ostojic_2019,Platkiewicz_Saccomano_McKenzie_English_Amarasingham_2021,Wei_Stevenson_2021}. The model fits shown above are based on a decomposition of the correlogram into a slow, baseline component and a fast, synaptic effect. However, to account for the dynamics of the putative presynaptic neuron, in some cases it may be necessary to incorporate the presynaptic autocorrelation into the model.

Due to the bursting or refractoriness of the presynaptic neuron, the influence of the synaptic effect on the correlogram may be spread across a wide range of intervals (Fig 5). A strong putative excitatory synapse from a presynaptic neuron that bursts, may have multiple peaks in the correlogram corresponding to the modes for bursts in the ISI distribution (Fig 5A and B). Similarly, a strong putative inhibitory synapse from a presynaptic neuron with refractoriness, may show a “release from inhibition” for intervals shortly before 0 (Fig 5C). Many putative synapses (see examples in Fig 1, 3, and 4) appear on a relatively constant background firing, but these cases (taken from the Allen Brain Institute – Visual Coding Neuropixels dataset) illustrate how presynaptic bursting and refractoriness can predictably influence the observed correlogram.

To account for this variation, here we use a convolutional model of the fast synaptic effect where a fast, fixed function (in this case, an alpha function) is convolved with the presynaptic autocorrelation (see Methods). Previous studies have noted the importance of the presynaptic autocorrelation in shaping the cross-correlogram \citep{Bryant_Marcos_Segundo_1973,Moore_Segundo_Perkel_Levitan_1970} and shown how deconvolution could be used to correct for these effects during detection, testing, and estimation \citep{Eggermont_Smith_Bowman_1993,Spivak_Levi_Sloin_Someck_Stark_2022}. Here we find that the same convolutional synaptic model (with a single estimated latency, time constant, and strength) can describe the cross-correlograms associated with subsets of presynaptic spikes. Here we consider isolated presynaptic spikes (those where the preceding/following presynaptic spikes are somewhat distant) and non-isolated presynaptic spikes (those with a nearby preceding/following presynaptic spike), and find that, after applying the convolution with the presynaptic auto-correlogram, a single synaptic model describes the overall correlogram, as well as the correlograms associated with isolated and non-isolated spikes (Fig 5).

\begin{figure}
\centering
\includegraphics[width=\textwidth*4/5]{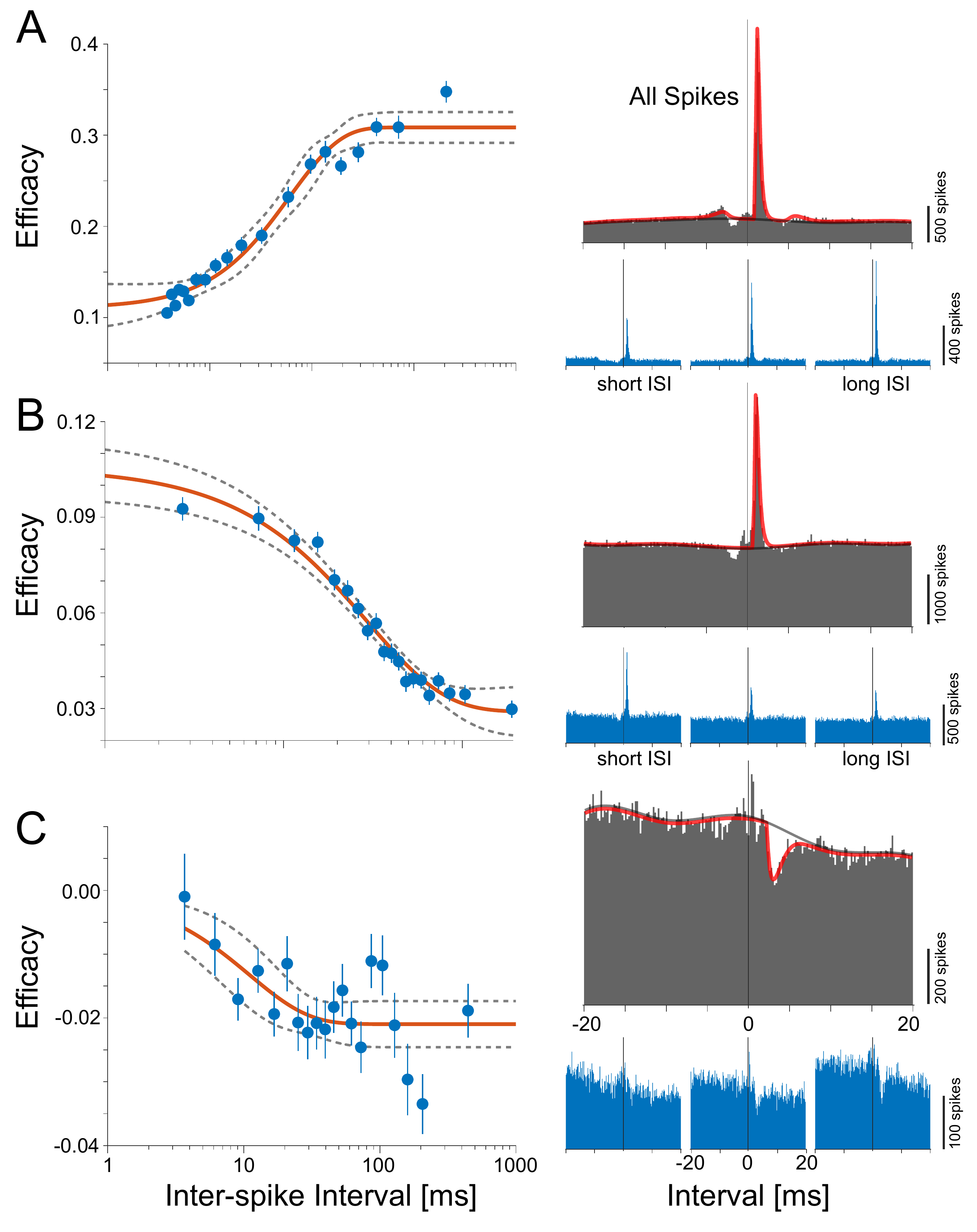}
\caption{Example putative synapses (Allen Institute Neuropixels) with parameters estimated using pairwise model-based methods. A) Putative excitatory CA1-CA1 connection (Session ID 719161530) with short-term synaptic depression. The cross-correlogram of postsynaptic spikes (gray) shows a transient, short latency peak following presynaptic spiking. The red curve denotes the model fit, and black curve denotes the baseline (without the synaptic effect). As the inter-spike interval preceding a presynaptic spike increases the efficacy increases, suggesting recovery from depression (left). The correlograms corresponding to presynaptic spikes short, medium, and long ISIs are shown in blue. B) Putative excitatory VISam-VISam connection (819701982) with short-term synaptic facilitation. C) Putative inhibitory VPM-VPM connection (719161530) with short-term synaptic depression.}
\end{figure}
This approach accounts for some types of variation in the cross-correlation, but it ignores the potential for short-term synaptic dynamics. The postsynaptic response to the first spike in a burst is not expected to be the same as the last spike in a burst due to short-term synaptic depression and facilitation \citep{Swadlow_Gusev_2001,Usrey_Reppas_Reid_1998}. Putative synapses can show patterns consistent with short-term synaptic depression, where synaptic effects for presynaptic spikes preceded by long ISIs are stronger than those for presynaptic spike preceded by short ISIs (Fig 6A), or consistent with short-term synaptic facilitation, where the effects are reversed (Fig 6B). Explanatory models of short-term synaptic plasticity that describe resource depletion and recovery as well as release probability and membrane potential integration can accurately describe these patterns \citep{English_McKenzie_Evans_Kim_Yoon_Buzsáki_2017}, and here we show several examples from the ABI dataset (Fig 6, see Methods for model details). The more detailed responses to presynaptic dynamics allow explanatory models to be tested and compared. Models fit to data from one stimulus or brain state can be tested to see if they are able to generalize and predict effects during other stimuli or brain states.

\subsection{Confounding by polysynaptic connections and common drive}
As previous authors have noted, polysynaptic connections and common drive can, in principle, confound the identification of monosynaptic connections \citep{Brody_1999,Gerstein_Perkel_1969,Knox_1981,Nykamp_2008}. Strong disynaptic effects through a hidden neuron (e.g. A→H→B) may lead to peaks or troughs in the cross-correlogram with short latencies and durations (between neurons A and B, Fig 7A), and strong common drive (H→A and H→B) can lead to similar synapse-like statistics when delays are coordinated (again, between neurons A and B, Fig 7B).

To understand how common this confounding is likely to be, we consider some approximate calculations of how efficacious polysynaptic transmission and common drive are in practice. On the one hand, there is reason to think the confounding may not be severe. For an excitatory disynaptic connection A→H→B with efficacies $e_AH>0$ and $e_HB>0$, the polysynaptic efficacy of A to C will be $\sim e_AH$ $e_HB$ assuming spike transmission is independent. Similarly, for the case where A and B receive common drive from H, the excess probability of A and B both spiking will be $\sim e_HA$ $e_HB$. For the range of monosynaptic efficacies observed in vivo on the order of $\sim 5\%$, disynaptic transmission or common drive would have an efficacy $\sim 0.25\%$. Additionally, only a subset of confounding effects will have latencies and durations likely to be confused for monosynaptic connections. The latencies and durations of polysynaptic effects grow linearly with additional connections, while the efficacy drops geometrically (Fig 6A). On the other hand, common drive must occur with specific delays to produce latencies consistent with monosynaptic transmission (Fig 7B). Multiple paths from A→B (Fig 7C) or multiple hidden inputs (Fig 7D) could act to increase the effective efficacy. However, to be comparable to a typical monosynaptic connection many such paths would be necessary, and these paths/inputs would need to have precisely coordinated latencies to prevent broadening of the cross-correlation. Imprecise mixtures of polysynaptic and common drive effects (Fig 7E) will generally not result in cross-correlograms that would be confused for a fast monosynaptic effect.

Cross-correlations between pairs of neurons have been observed and described on multiple timescales, and previous work has aimed to distinguish between common input, stimulus-driven, and emergent synchrony \citep{Ostojic_Brunel_Hakim_2009,Usrey_Reid_1999}. Sensory-driven and spontaneous activity in sensory cortex can also occur with near-synchronous spiking \citep{Bair_Zohary_Newsome_2001,Kohn_Smith_2005,Swadlow_Beloozerova_Sirota_1998,Toyama_Kimura_Tanaka_1981}, and similar fast correlations have been observed between hippocampal \citep{Diba_Amarasingham_Mizuseki_Buzsáki_2014} and thalamic neurons \citep{Alonso_Usrey_Reid_1996}.  In most cases, synchronous common input does not have a clear directionality and is centered at zero latency (see examples from ABI-715093703 in Fig 4D and S1), but short duration cross-correlations with non-zero latency can still have ambiguous origins (e.g. \cite{Atencio_Shen_Schreiner_2016}). Disynaptic effects have also been studied directly in the projections from retina to visual cortex via the LGN \citep{Kara_Reid_2003}. Here Reid and colleagues found that cross-correlations for disynaptic effects resemble the convolution of retinogeniculate and geniculocortical synaptic effects \citep{Reid_2001}. The duration of the disynaptic effect is longer and the efficacy is smaller than for each of the individual synapses, but it may be fast/strong enough that, without accounting for anatomy, the disynaptic effect could be mistaken for a monosynaptic effect.

Altogether, these observations suggest how polysynaptic effects and common drive are at the same time 1) relatively unlikely to be mistaken for monosynaptic effects, on average, and 2) possible confounds in concrete cases. The short duration and non-zero latency of monosynaptic transmission provides some protection against confounding by slow effects, but the overall cross-correlogram alone cannot uniquely differentiate between monosynaptic effects and other fast dependencies between pre- and postsynaptic spikes. The circumstantial evidence discussed above, however, may provide an additional layer of defense against mistaking polysynaptic effects and common drive for monosynaptic transmission. In cases where cross-correlations have a short-latency, transient peak or trough due to confounding, other evidence may argue against the presence of a monosynaptic connection: Dale’s Law may be violated, spike waveforms may be inconsistent with observed “excitation” or “inhibition”, or effects may be improbable given the anatomical locations of the putative pre- and postsynaptic neurons.

\begin{figure}[bt]
\centering
\includegraphics[width=\textwidth*3/4]{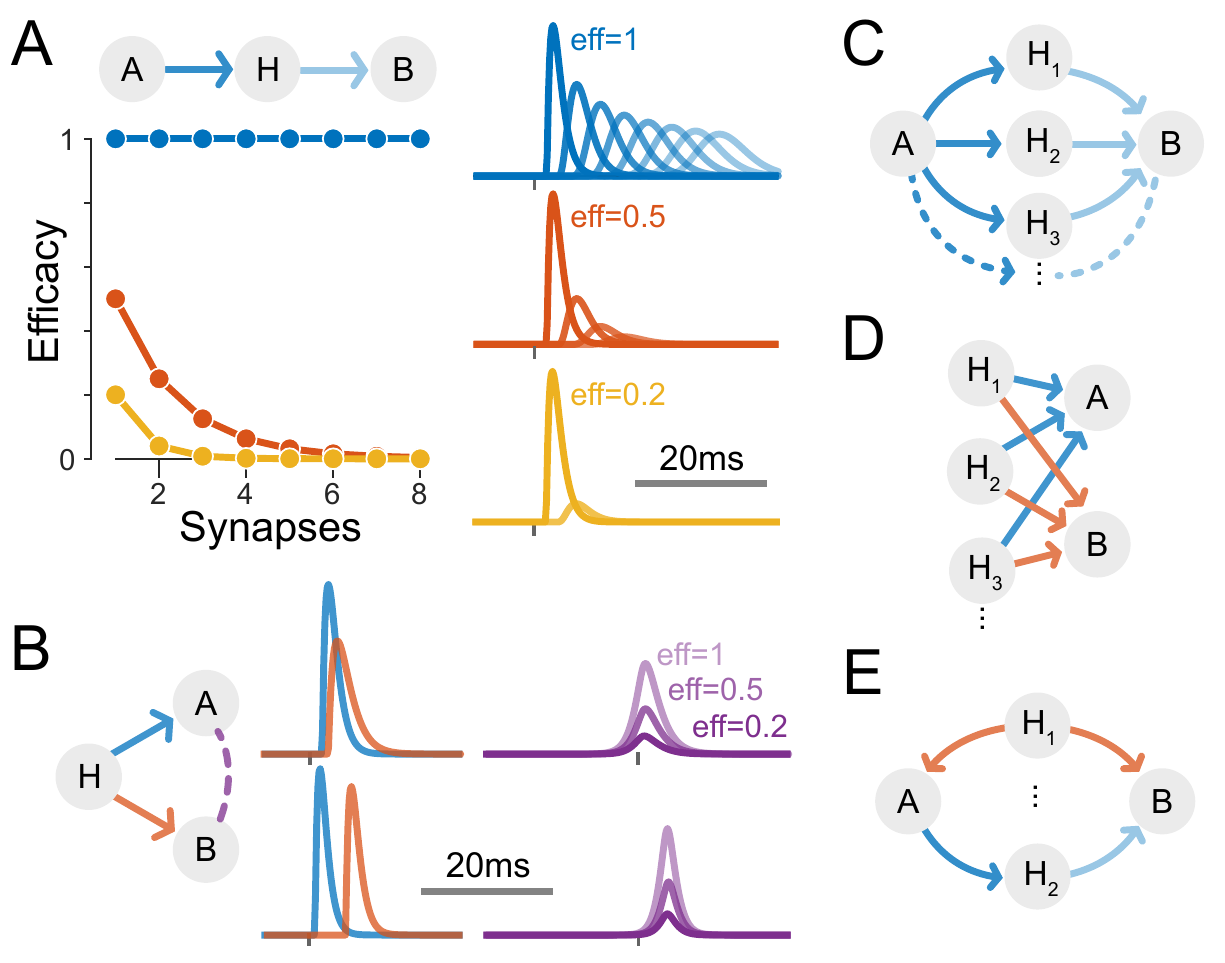}
\caption{Spurious correlations can occur due to polysynaptic chains and/or common input. However, in order to be mistaken for monosynaptic effects, these confounders need to have specific structure. A) Since the efficacy of single synapses tends to be much less than 1, the strength of single polysynaptic chains decays rapidly. The latency and time constant of the effect also grows with the number of connections. Thus, only chains with strong connections and relatively few links are likely to be mistaken for direct, monosynaptic connections. B) Common drive from hidden neurons can also create spurious correlations. Again, since the efficacy of single connections is much less than 1, these effects will tend to be weaker and somewhat less temporally precise. Additionally, unless the latencies of the connections are substantially different, the effect of common drive will be centered at zero delay and, thus, be less likely to be mistaken for a monosynaptic connection. Although the effects of individual polysynaptic chains or common drive from an individual input may be weak, multiple chains (C), multiple drivers (D), or a combination of the two (E) can create spurious correlations comparable in strength and timing to monosynaptic connections, albeit with even less temporal precision than single inputs/chains.}
\end{figure}

\begin{figure}
\centering
\includegraphics[width=\textwidth]{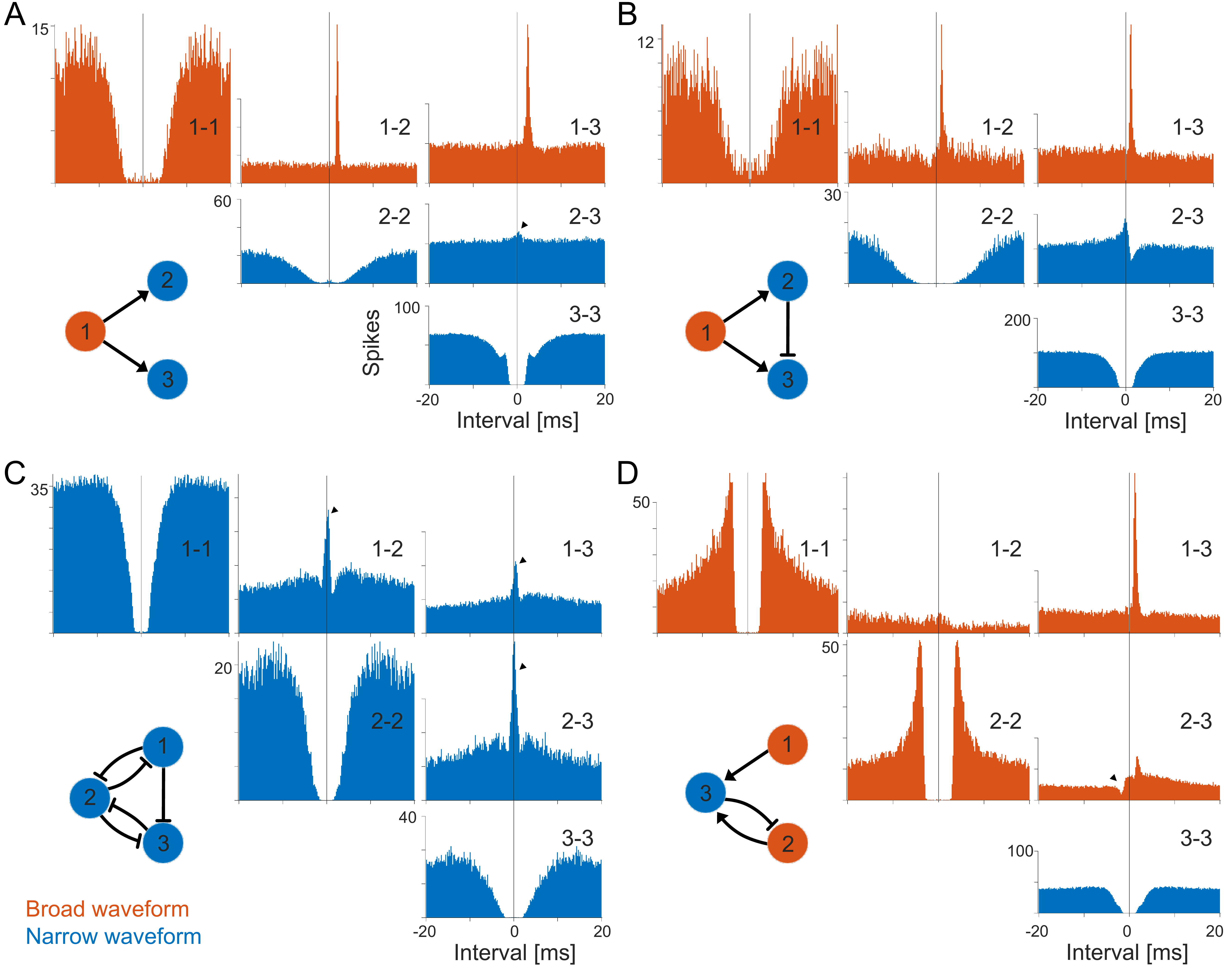}
\caption{Examples of putative subnetworks involving three single units. A) Auto-correlograms and cross-correlograms for putative common drive from a single unit with a broad waveform to two units with narrow waveforms (VIS, Session ID 744228101). Triangle denotes the potential, spurious correlation between the two narrow waveform units due to their shared excitatory input. B) Putative feed-forward inhibition (VISal, 760693773), C) Putative inhibitory subnetwork (VISrl, 746083955). Triangles denote the synchronous activity of all three units. D) Putative microcircuit with a reciprocal connection (grey, 742951821). Triangle denotes the putative reciprocal inhibition. Correlogram columns within each panel correspond to the same putative postsynaptic neurons and have the same axes. Putative presynaptic units whose spike waveforms were classified as broad are colored in red, while those classified as narrow are shown in blue.}
\end{figure}

\subsection{From putative synapses to putative microcircuits}
Careful consideration of confounding is necessary whenever there are unobserved influences on postsynaptic spiking \citep{Nykamp_2007}, but, as the scale of multi-electrode recordings increases, observed disynaptic effects (e.g. A→B→C) and observed diverging outputs from a common presynaptic neuron (e.g. A→B and A→ C) may also be studied more directly, alongside other putative microcircuit motifs. In the ABI datasets, there are many examples of neurons that appear to be, not just synaptically connected as an isolated pair, but connected in larger subnetworks (Fig 8). In one example (Fig 8A), a putative excitatory presynaptic neuron (with a broad waveform) appears to project to two putative inhibitory neurons (with narrow waveforms). The weak, near-synchronous correlation between the two inhibitory neurons may be due to their common input. More complex patterns are also common – for instance, putative feed-forward inhibition (Fig 8B), putative inhibitory subnetworks (Fig 8C), and putative networks with reciprocal connections (Fig 8D).

Motifs are hypothesized to be building blocks of information processing \citep{Braganza_Beck_2018,Wang_Yang_2018}, and there are widespread similarities in microcircuits across species and brain regions \citep{Silberberg_Grillner_LeBeau_Maex_Markram_2005}. Although motifs have been observed in previous studies of putative synapses \citep{Barthó_Hirase_Monconduit_Zugaro_Harris_Buzsáki_2004,Fujisawa_Amarasingham_Harrison_Buzsáki_2008}, description and modeling of these motifs during behavior is still in the early stages. Despite the unmodeled influence of unobserved inputs, explanatory models allow direct predictions of the motif effects. Feed-forward inhibition in the cerebellum, for instance, has been well-modeled by the combination of putative synaptic effects \citep{Blot_de_Solages_Ostojic_Szapiro_Hakim_Léna_2016}. As large-scale spike recordings increase in scale, searching for and modeling putative synaptic effects may complement intracellular studies of these motifs e.g. \citep{Pouille_Scanziani_2001,Wehr_Zador_2003} and allow predictions based on in vitro studies to be tested in behaving animals.

\section{Discussion}
Here we have argued that synaptic connections can reasonably be studied from large-scale spike recordings. As the scale of spike recordings increases these connections will inevitably occur in the data, and there is strong evidence that at least some synapses can be detected from spikes alone under realistic physiological conditions. We show how putative connections detected during observation are often supported by multiple lines of circumstantial evidence, and, although confounding is still a concern, models built with synaptic effects can provide compelling explanations for the detailed structure of pairwise spike statistics. These models do not give unbiased estimates of verified, causal synaptic effects when fit to observational data, but they do provide testable descriptions of the data with synaptic parameters, dynamics, or microcircuit structure.

Here we have particularly emphasized the utility of detailed explanatory models and intersecting evidence from sources other than the overall correlogram for interpreting putative synapses. Although, the overall cross-correlogram between the spikes of neuron A and neuron B provides “merely” observational evidence for the potential presence of a synapse, our confidence in the presence of a synapse can be strengthened or weakened by integrating prior physiological knowledge and by testing whether explanatory models make accurate predictions. We have focused here largely on the qualitative role that circumstantial evidence can play in interpreting spike correlations, but, in some cases, it may be useful to quantitatively integrate circumstantial evidence into the detection and modeling process, for instance, using Bayesian methods \citep{Ren_Ito_Hafizi_Beggs_Stevenson_2020}. Here we have reviewed how information about network structure (e.g. Dale’s law), anatomy, and spike waveforms can be used to assess whether candidate correlograms are consistent with synaptic effects. And we have shown how detailed consideration of the latency and duration of synaptic effects as well as modeling of presynaptic dynamics can further alter our confidence. In observational studies, this “multi-phasic” approach to evaluating the evidence \citep{Cochran_1965} has been called “Cochran’s crossword” \citep{Rosenbaum_2015}. Although we may be uncertain about a specific answer (the presence of a synapse) given a single clue (the overall cross-correlogram), intersecting clues can, like a crossword puzzle, support or contradict our belief that two neurons may be connected.
\\

Many methods have been developed for measuring statistical dependencies between pairs of neurons \citep{Magrans_de_Abril_Yoshimoto_Doya_2018}. Although the correlogram itself, and other measures such as Granger causality and transfer entropy, can provide important descriptions of dependencies between pairs of spike trains, the scientific goal of the explanatory models described here is somewhat distinct. Having a detailed explanatory model \citep{Hand_2019,Shmueli_2010} that predicts postsynaptic responses allows measurement but also allows multiple explanations to be compared and falsified. Comparisons beyond the data, for instance, to anatomy \citep{Gerhard_Kispersky_Gutierrez_Marder_Kramer_Eden_2013} or across stimuli \citep{Ghanbari_Ren_Keine_Stoelzel_Englitz_Swadlow_Stevenson_2020}, play an essential role in the interpretation of the results. New models that incorporate either more biophysical detail, such as membrane dynamics \citep{Ladenbauer_McKenzie_English_Hagens_Ostojic_2019,Platkiewicz_Saccomano_McKenzie_English_Amarasingham_2021}, or more complex forms of spike transmission and plasticity \citep{Linderman_Stock_Adams_2014,Song_Robinson_Berger_2018,Wei_Stevenson_2021} are actively being developed, and these models may provide even more accurate descriptions and predictions of postsynaptic spiking. Accurate descriptions of the fluctuations in postsynaptic excitability and adaptation (c.f. \cite{Kobayashi_Tsubo_Shinomoto_2009}) may be especially important, since these effects often occur on similar timescales to short- and long-term plasticity.

When the standard of evidence is strict, errors and biases in the identification of putative synapses are not qualitatively worse than errors and biases in spike sorting \citep{Ventura_2009}, firing rate estimation \citep{Amarasingham_Geman_Harrison_2015}, cell type identification \citep{Lee_Balasubramanian_Tsolias_Anakwe_Medalla_Shenoy_Chandrasekaran_2021,Trainito_von_Nicolai_Miller_Siegel_2019}, sampling of units, or the targeting of brain regions. However, for studying synaptic effects, one major bias is in the strength of detected connections – strong synapses between neurons with high firing rates are more likely to be detected than weak connections between neurons with low firing rates. False positives may occur if common input is mistaken for a synaptic effect, but there are likely to be many, many false negatives. The synapses that are detected, sometimes referred to as the “billionaires” (Swadlow, personal communication), are likely to be unrepresentative of typical synaptic effects and many not accurately reflect typical microcircuit operation. Even if we can be confident that a putative synapse is likely to be a genuine synapse, it is important to keep in mind that we may be studying the most impactful synapses and not the typical synapses. Here we have focused exclusively on efficacy as a measure of impact. However, many previous studies emphasize contribution rather than or in addition to efficacy (c.f. \cite{Miller_Escabí_Read_Schreiner_2001}). Note that for excitatory synapses, contribution can be rather directly calculated by scaling the efficacy by $N_{pre}/N_{post}$ where $N$ denotes total spikes for the pre- or postsynaptic neuron. We focus on efficacy here, since contribution is ill-defined for inhibitory synapses.

Without experimental interventions, there will always be uncertainty about whether a specific pair of neurons is genuinely synaptically connected, and unobserved confounds can generally lead to biases in the estimated dependencies between neurons, just as they do for other estimates \citep{Stevenson_2018}. Nonetheless, here we argue that the problem is, perhaps, not as dire, as several recent results have suggested \citep{Das_Fiete_2020,Liang_Brinkman_2023,Mehler_Kording_2020}. In the case of putative synapses, circumstantial evidence and explanatory models can act to mitigate concerns about confounding, and focusing on these effects allows synaptic transmission to be studied in intact neural systems. However, efforts to identify putative synapses should be distinguished from other efforts to describe functional connectivity \citep{Reid_Headley_Mill_Sanchez-Romero_Uddin_Marinazzo_Lurie_Valdés-Sosa_Hanson_Biswal_et_al._2019,Stevenson_London_Oby_Sachs_Reimer_Englitz_David_Shamma_Blanche_Mizuseki_et_al._2012} that are at least an order of magnitude slower and denser.

\section{Methods}

Code for the results shown here is available at \href{https://github.com/stevensonlab/syngalong_v0-1}{https://github.com/stevensonlab/syngalong$\_$v0-1}

\subsection{Simplified power analysis}
Consider a hypothesis test of a synaptic connection with efficacy $e$ given $N$ presynaptic spikes and known probability $p$ for the postsynaptic neuron firing by chance within a fixed detection window following each presynaptic spike. We wish to decide between two hypotheses: $H_0:e=0$ versus $H_A:e \neq 0$. Assuming that the observations are independent, the null distribution for the number of postsynaptic spikes in the detection window is given by $y \sim Binomial(N,p)$ and alternative distribution is $y \sim Binomial(N,p+e)$. Using the normal approximation to the binomial distribution $Binomial(N,p) \approx Normal(Np,\sqrt{Np(1-p)}$, with confidence level $\alpha$, the approximate power $(1-\beta)$ is given by
\begin{equation*}
1-\beta=1+\Phi \left( \frac{\sqrt{Ne^2}+\sqrt{p(1-p)}) z_{\alpha/2}}{\sqrt{(p+e)(1-p-e)}} \right) -\Phi \left( \frac{\sqrt{Ne^2}+\sqrt{p(1-p)} z_{1-\alpha/2}}{\sqrt{(p+e)(1-p-e)}} \right)
\end{equation*}
where $\Phi(\cdot)$ denotes the CDF for the standard normal distribution and $z_\cdot$ denotes the quantiles of the standard normal distribution (e.g. $z_0.975=1.96$). We can approximately solve for the efficacy required to achieve a given power using a first-order Taylor expansion around e=0. We have two solutions, one for excitatory connections and one for inhibitory
\begin{align*}
e^+ &\approx \frac{p(1-p) \left( z_{1-\alpha/2}-z_\beta \right) }{\sqrt{Np(1-p)}-(p-\frac{1}{2}) z_\beta }\\
e^- &\approx \frac{p(1-p) \left( z_{\alpha/2}-z_{1-\beta} \right)}{\sqrt{Np(1-p)})-(p-\frac{1}{2}) z_{1-\beta}}
\end{align*}
These are upper-bounded and equal when $p=0.5$, in which case $e_{p=0.5} \approx (z_{1-\alpha/2}-z_\beta)/(2\sqrt{N})$. For the commonly used values of $\alpha=0.05$ and $1-\beta=0.8$ this gives $e_{p=0.5} \approx 1.4/\sqrt{N}$.

In practice, there are major caveats to this analysis: 1) it ignores the problem of how to define the time window of interest, 2) it ignores the fact that the shape/level of the baseline p is unknown, 3) it ignores the dependencies that exist between post-synaptic spikes, and 4) it ignores the complex, overdetermined relationships between biophysical parameters and efficacy. Many methods for detecting synapses from spikes have aimed to address Caveats 1, 2, and 3. However, with these computational and model-based methods, simulations are necessary to provide estimates of detection accuracy in specific settings. Rather than an exact power analysis, these results often focus on generating estimates of false positive rates ($\hat{\alpha}$) and false negative rates ($\hat{\beta}$). Caveat 4, is an important consideration for linking results with extracellular recordings to the many previous intracellular studies of synaptic transmission \citep{Ladenbauer_McKenzie_English_Hagens_Ostojic_2019,Platkiewicz_Saccomano_McKenzie_English_Amarasingham_2021}. Efficacy cannot be directly inferred from post-synaptic currents or potentials (PSC/PSPs) but depends on the nonlinear transformation between membrane potential and spikes. Differences in receptor time constants (e.g. AMPA vs NMDA or GABAA vs GABAB) and the presence of additional currents, mean that two synapses with similar PSC/PSPs may have quite different efficacies and two synapses with the same efficacies can have quite different biophysical origins. Considering efficacy directly prevents us from making precise conclusions about synaptic biophysics but allows us to derive approximate limits for detecting synaptic connections.

\subsection{Data}
Here we examine two datasets as case studies: 1) an in vitro recording of spontaneous activity from an organotypic slice culture of somatosensory cortex, and 2) an in vivo recording from multiple Neuropixels arrays in an awake mouse.

Slice data were previously collected and are described in detail in  \cite{Ito_Yeh_Hiolski_Rydygier_Gunning_Hottowy_Timme_Litke_Beggs_2014}. Briefly, recordings were made using a large-scale multielectrode array (1 x 2 mm total area with 512 electrodes spaced 60 $\mu m$ apart on a hexagonal lattice). Single units were visually identified and sorted using multichannel waveforms from each electrode and its six neighbors. Data are available via the Collaborative Research in Computational Neuroscience (CRNCS) Data Sharing Initiative: http://dx.doi.org/10.6080/K07D2S2F.

Neuropixels recordings from the Allen Institute for Brain Science – Visual Coding dataset were previously collected and described in detail in \cite{Siegle_Jia_Durand_Gale_Bennett_Graddis_Heller_Ramirez_Choi_Luviano_et_al._2021}. Briefly, head-fixed mice viewed a standardized set of visual stimuli (including Gabor patches, full-field drifting gratings, moving dots, and natural images and movies) while they were free to run on a wheel. Data are available from the Allen Brain Institute (https://portal.brain-map.org/explore/circuits).

\subsection{Detecting putative monosynaptic connections}
We detect putative connections based on the cross-correlograms between pairs of neurons. Several approaches for detecting putative synapses have been developed based on hypothesis testing \citep{Amarasingham_Harrison_Hatsopoulos_Geman_2012,Fujisawa_Amarasingham_Harrison_Buzsáki_2008} or model comparison \citep{Kobayashi_Kurita_Kurth_Kitano_Mizuseki_Diesmann_Richmond_Shinomoto_2019,Ren_Ito_Hafizi_Beggs_Stevenson_2020}. Here, since we are focused on the modeling and interpretation of putative synapses rather than detection per se, we detect putative monosynaptic connections using a computationally efficient, approximate hypothesis test based on the jitter method. Namely, we compare the observed cross-correlogram to a version of the cross-correlogram that has been smoothed and evaluate the probability of the observed spike count in each bin under a null distribution.

For the exact jitter method, a null distribution for $c(\tau)$ is generated based on the empirical correlograms when observed spike times are jittered. Here we evaluate p-values $c(\tau)$ based on a null distribution $Binomial(s(\tau),p(\tau))$ that assumes individual counts in the correlogram are conditionally independent given $\tau$. The trials $s$ and probability $p$ associated with each bin of the correlogram are generated by a “transport” function $g(t)$. Conceptually, $s(\tau)$ reflects the number of spikes that could be jittered from neighboring bins into bin $\tau$
\begin{equation*}
s(\tau)=\sum_t \mathbb{1}_{|\tau-t|<m} c(\tau-t)g(t)
\end{equation*}
and $p(\tau)$ reflects the probability of a spike being transported into the given bin
\begin{equation*}
p(\tau)=1/\sum_t \mathbb{1}_{|\tau-t|<m}g(t)
\end{equation*}
Here we use a Gaussian window $g(t)=exp(-\frac{t^2}{2\sigma^2})$ with a standard deviation of $\sigma=2 ms$. Note that when $\sigma$ is bigger than the binsize, $\sum_t g(t) >1$ and $s(\tau)>c(\tau)$. More spikes could land in bin $\tau$ than actually do. When the correlogram is calculated over a fixed, symmetric interval $-m\ge \tau \ge m$ and $\sigma<<m$, $p$ is approximately constant but increases to account for the edges of the correlogram. Since this approach is based on simply smoothing the correlogram, it is substantially faster than the simulation-based jitter method, and effectively detects any variation in the correlogram that is fast compared to the timescale of the smoothing.

When there are strong dependencies between individual counts in different bins this approximation will not be as accurate as the jitter method. For example, if the pre-synaptic neuron bursts nearby post-synaptic spikes will appear as non-independent counts in the correlogram. However, in practice, we find that the independence assumption detects reasonable connections with substantial computational savings.

Testing B bins within $1 ms<\tau<25 ms$ of each correlogram with C neurons results in $O(BC^2)$ hypothesis tests. To correct for multiple comparisons, we use a False Discovery Rate method (Benjamini-Hochberg procedure) with $\alpha=0.1$ and consider pairs with at least two statistically significant time bins to be a putative connection. Increasing $\alpha$ would result in more pairs being detected with greater numbers of weaker connections and more false positives \citep{Ren_Ito_Hafizi_Beggs_Stevenson_2020}.

In some cases, overlapping spike waveforms from nearby neurons (spike shadowing) leads to misestimation of $c(\tau)$ near $\tau=0$ \citep{Pillow_Shlens_Chichilnisky_Simoncelli_2013,Quirk_Wilson_1999}. Here we detect these cases using a spike shadowing index 
\begin{equation*}
SSI = min \left( \frac{ \left| \bar{c}_{|\tau|<t_w}-\bar{c}_{-2t_w<\tau<-t_w} \right|}{\bar{c}_{-2t_w<\tau<-t_w}}, \frac{\left| \bar{c}_{|\tau|<t_w}-\bar{c}_{t_w<\tau<2t_w} \right|}{\bar{c}_{t_w<\tau<2t_w}} \right)
\end{equation*}

This index compares the average observed count (denoted $\bar\cdot$) in the window of potential overlap $|\tau|<t_w$ to the neighboring regions of the correlogram on either side and is large when there is abrupt variation in the correlogram near $\tau=0$. Here we set $t_w=1 ms$ and exclude pairs of neurons with $SSI>0.4$ (7$\%$ of pairs, in Fig. 1). Recent spike sorting methods that explicitly consider overlapping waveforms \citep{Pachitariu_Steinmetz_Kadir_Carandini_Harris_2016} typically have much less spike shadowing and may correct for sorting errors due to changing spike shape. For putative synapses between units on different arrays or on electrodes that are far apart, spike shadowing is not a problem and $SSI$ is often near 0.

\subsection{Modeling synaptic effects in the cross-correlogram}

Although the hypothesis-test based approach is effective for rapidly detecting connections, it only looks for differences between fast and slow structure in the correlogram and does not provide a generative model for synaptic transmission. In order to assess whether putative connections are excitatory or inhibitory, measure their latencies, and efficacies we fit an extended GLM, similar to \cite{Kobayashi_Kurita_Kurth_Kitano_Mizuseki_Diesmann_Richmond_Shinomoto_2019, Ren_Ito_Hafizi_Beggs_Stevenson_2020}.

Namely, we model the counts in the cross-correlogram using Binomial regression with
\begin{equation*}
p_\tau=f \left( \beta_0 + X_\tau \beta + a_\tau * w\alpha(t) \right)    
\end{equation*}
where $X_\tau$ denotes a set of fixed basis functions weighted by coefficients $\beta$ to describe slow fluctuations in the baseline correlogram, while the fast synaptic effect is modeled by convolving an alpha function $\alpha(t)=\frac{t-t_0}{\tau_\alpha} exp\left( -\frac{t-t_0}{\tau_\alpha} \right)$ with the auto-correlation of the presynaptic neuron $a_\tau$ with synaptic weight $w$.

Here $t_0$ and $\tau_\alpha$ denote the latency and time constant for the synaptic effect, respectively, and the convolution allows temporal dependencies in the presynaptic firing (e.g. burstiness) to be taken into account. Here, to model the slow fluctuations in the correlogram we use a set of cubic B-spline basis functions with equally spaced knots (between 3-6 for the fits shown here). The combination of the slow effect and fast synaptic effect is then passed through a logistic nonlinearity $f(x)=1/(1+e^{-x})$, and the number of counts observed in the correlogram is assumed to be distributed following $Binomial(N,p_\tau)$ where $N$ is the number of presynaptic spikes.

The parameters $\theta=\left\{ \beta_0,\beta,w,t_0,\tau_\alpha \right\}$ are optimized by maximizing a penalized log-likelihood
\begin{equation*}
log p(c|\theta) \propto \sum_\tau \left[ c_{\tau} log(p_\tau)+(N-c_\tau)log(1-p_\tau) \right] -\eta_\beta ||\beta||^2-\eta_w w^2-\eta_\tau \tau_\alpha
\end{equation*}

using gradient descent, and the hyperparameters $\eta=\left\{ \eta_\beta,\eta_w,\eta_\tau \right\}$ act to regularize the basis function coefficients, synaptic amplitude, and synaptic time constant, respectively. Due to the parameterization of the alpha function, the log likelihood is not concave, and we use random restarts to find the best solution. The synaptic latency and time constant are both constrained to be positive by log-transforming during optimization. After optimization, the efficacy is estimated by
\begin{equation*}
e=\sum_{\tau \in \alpha>0.01} (p_\tau-f(\beta_0+X_\tau beta))     
\end{equation*}

To account for asymmetry in the effect of the auto-correlogram related responses we, in some cases (e.g. Fig 5A), separate the auto-correlogram into two segments and add the additional parameter $v$
\begin{equation*}
p_\tau=f \left( \beta_0 + X_\tau \beta + a_{\tau<0}*v \alpha(t) + a_{\tau \ge 0}*w \alpha(t) \right)
\end{equation*}

The log-likelihood itself can be used for detecting putative synapses \citep{Ren_Ito_Hafizi_Beggs_Stevenson_2020}, but here we focus on the model fits of already suspected synapses, using pseudo-$R^2$ to assess goodness-of-fit
\begin{equation*}
R^2=\frac{L_\theta-L_0}{L_s-L_0}
\end{equation*}
where $L_\theta$ denotes the (unpenalized) log-likelihood for the fit model, $L_0$ the log-likelihood for the null model (with constant $p=\bar{c}/N$), and $L_s$ the log-likelihood for the saturated model (with $p(\tau)=c(\tau)$).

\subsection{Modeling short-term synaptic plasticity}
Following \citep{English_McKenzie_Evans_Kim_Yoon_Buzsáki_2017} we fit a nonlinear model for the efficacy as a function of the presynaptic inter-spike interval $\Delta t$...
\begin{align*}
S(\Delta t) &= U exp \left( -\frac{\Delta t}{\tau_s} \right)\\
D(\Delta t) &= 1-U exp \left( -\frac{\Delta t}{\tau_d} \right)\\
F(\Delta t) &= U+\left[ U(1-U) \right] exp \left( -\frac{\Delta t}{\tau_f} \right)\\
\hat{e}(\Delta t) &= A \left[ S(\Delta t)+F(\Delta t)D(\Delta t) \right] 
\end{align*}

Where $S(\cdot)$, $D(\cdot)$, $F(\cdot)$ denote effects for synaptic integration, short-term synaptic depression, and short-term synaptic facilitation, respectively. $A$ denotes the amplitude of the synaptic effect, $U$ denotes the release probability, and $\tau_\cdot$ denotes the time constants of the three effects. The model here corresponds to the predictions of the Tsodyks-Markram model for a pair of presynaptic spikes that starts from a fully recovered state \citep{Tsodyks_Pawelzik_Markram_1998}. We fit the parameters $\theta = \left\{ A, U, \tau_s, \tau_d, \tau_f \right\} $ by minimizing the sum squared error $\sum_{\Delta t} \left( e(\Delta t)-\hat{e}(\Delta t) \right)^2$ between the model predicted ISI-dependency $\hat{e}(\Delta t) $ and the observed ISI-dependency $e(\Delta t)$. Here $e(\Delta t)$ is estimated by fitting the correlograms of subsets of presynaptic spikes that have specific quantiles of presynaptic ISIs $\Delta t$. We use the model above to first estimate the synaptic latency $t_0$ and time constant $\tau_\alpha$ for the overall correlogram, but reoptimize $w$ based on the subsets of presynaptic and their corresponding autocorrelograms. For the short-term plasticity model we use random restarts to avoid local minima, and, during optimization, the parameters are transformed $\theta'=\left\{ A,log(\frac{U}{1-U}),log(\tau_s), log(\tau_d), log(\tau_f) \right\} $ to constrain $0<U<1$ and $\tau_\cdot>0$.

\section*{acknowledgements}
This material is based upon work supported by the National Science Foundation under Grant 1651396. Thanks to Harvey Swadlow, Monty Escabí, Abhijith Mankili, Abed Ghanbari, and Naixin Ren for helpful discussions. Thanks to Shinya Ito, Alan Litke, and John Beggs and to the Allen Institute for Brain Science for sharing their datasets and for supporting open science.

%\section*{conflict of interest}
%fasdfasdfasdf
%\printendnotes

% Submissions are not required to reflect the precise reference formatting of the journal (use of italics, bold etc.), however it is important that all key elements of each reference are included.
\bibliography{sample}

%\begin{biography}[example-image-1x1]{A.~One}
%Please check with the journal's author guidelines whether author biographies are required. They are usually only included for review-type articles, and typically require photos and brief biographies (up to 75 words) for each author.
%\bigskip
%\bigskip
%\end{biography}

%\graphicalabstract{example-image-1x1}{Please check the journal's author guildines for whether a graphical abstract, key points, new findings, or other items are required for display in the Table of Contents.}

\begin{figure}[bt]
\renewcommand{\thefigure}{S1}
\centering
\includegraphics[width=\textwidth]{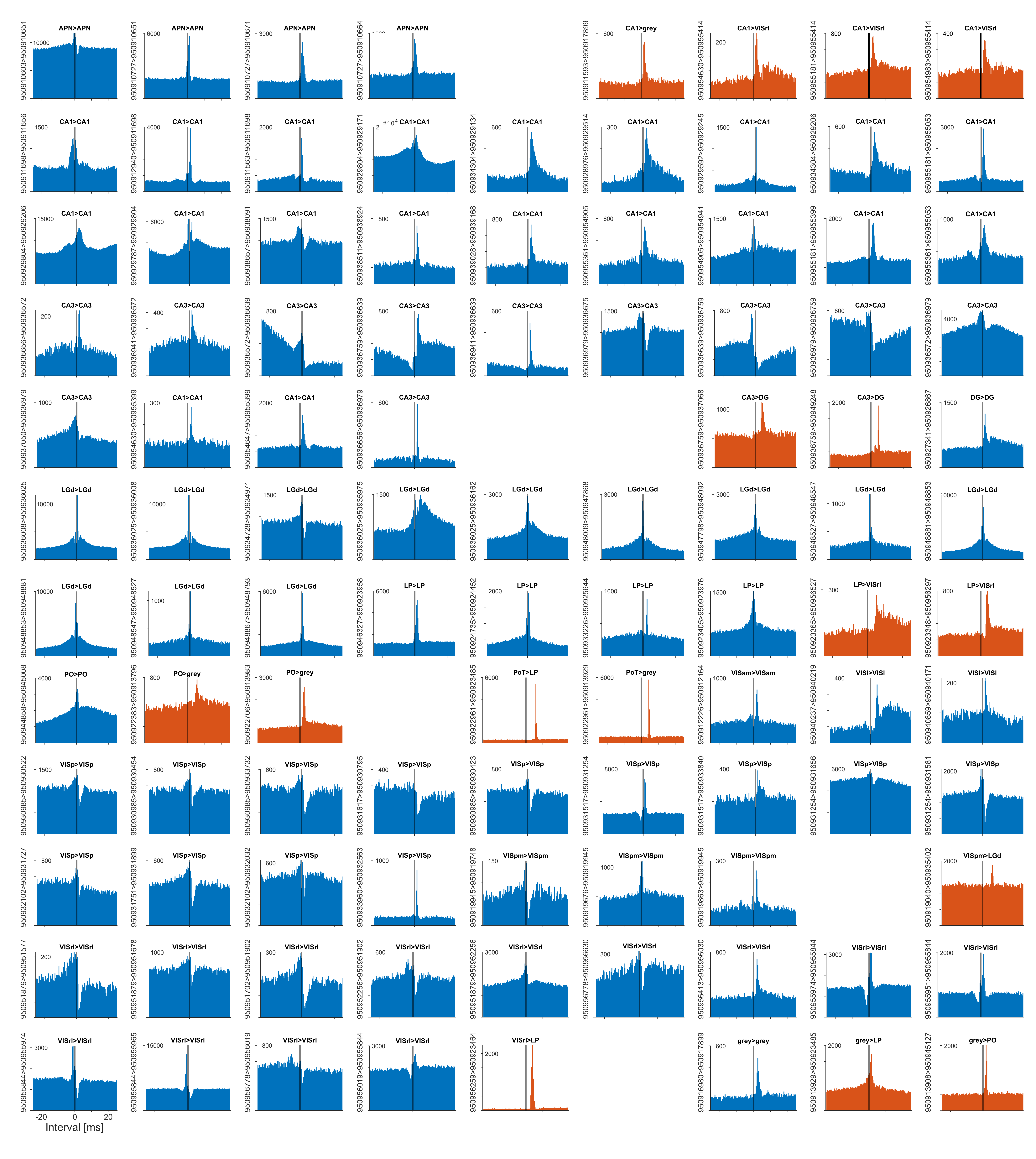}
\caption{Correlograms for $n=102$ pairs detected from ABI-715093703 screened for short latency and time constant and for high pseudo-$R^2$. Blue correlograms denote intra-area pairs, red correlograms denote pairs between areas. Labels correspond to the unique unit ids. 950948527>950948547, for instance, denotes a correlogram calculated “from” unit id 950948527 “to” unit id 950948547.}
\end{figure}

\end{document}